\RequirePackage{fix-cm,amsmath}
\documentclass[smallextended]{svjour3}       % onecolumn (second format)
\usepackage{graphicx}
\usepackage[utf8]{inputenc}
\usepackage[english]{babel}
\usepackage{wrapfig}
\usepackage[T1]{fontenc}
\usepackage{mathptmx}
\usepackage[english]{babel}
\usepackage{amsmath}
\usepackage{mathtools}
\usepackage{amsfonts}
\usepackage{amssymb}
\usepackage{makeidx}
\usepackage{scalerel}
\usepackage{hyperref}
\hypersetup{
    colorlinks=true,
    linkcolor=black,
    urlcolor=blue,
    citecolor=black
}
\usepackage[round]{natbib}
\usepackage[left]{lineno}
\usepackage{comment}
\usepackage{placeins}
\usepackage{xfrac}
\usepackage{color}
\usepackage{float}
\usepackage{caption}
\usepackage{subcaption}
\usepackage{chngcntr}
\usepackage{setspace}
\usepackage{array}
\usepackage{lscape}
\usepackage[labelsep=space]{caption}
 \setcitestyle{aysep={}} 
\counterwithin{figure}{section}
\counterwithin{table}{section}

\usepackage{todonotes}

\newcommand{\orcid}[1]{\href{https://orcid.org/#1}{\includegraphics[width=8pt]{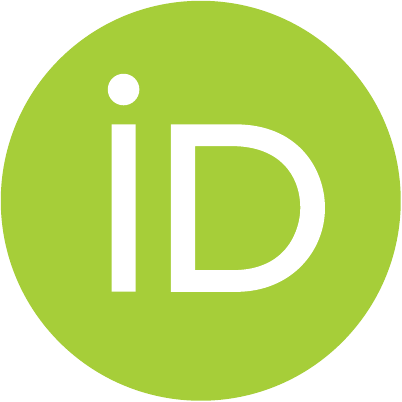}}}
\begin{document}

\title{Influence of Sodium Inward Current on Dynamical Behaviour of Modified Morris-Lecar Model}%\thanks{Grants or other notes



\author{H. O. Fatoyinbo  \orcid{0000-0002-6036-2957} \and S. S. Muni   \and A. Abidemi  \orcid{0000-0003-1960-0658}} %etc.


\institute{   H. O. Fatoyinbo \\
              School of Fundamental Sciences, Massey University, 
              Palmerston North, New Zealand\\
             \email{h.fatoyinbo@massey.ac.nz}
             \\ S. S. Muni \\ School of Fundamental Sciences, Massey University, 
              Palmerston North, New Zealand\\
             \email{s.muni@massey.ac.nz}
             \\
              A. Abidemi 
              \\ Department of Mathematical Sciences, Federal University of Technology, Akure\\
              \email{aabidemi@futa.edu.ng}
}

\date{Received: date / Accepted: date}

\maketitle

\begin{abstract}
This paper presents a modified Morris-Lecar model by incorporating the sodium inward current. The dynamical behaviour of the model in response to key parameters is investigated. The model exhibits various excitability properties as the values of parameters are varied. We have examined the effects of changes in maximum ion conductances and external current on the dynamics of the membrane potential. A detailed numerical bifurcation analysis is conducted. The bifurcation structures obtained in this study are not present in existing bifurcation studies of original Morris-Lecar model. The results in this study provides the interpretation of electrical activity in excitable cells and a platform for further study.

%to explore the influence of the maximum conductances of ion channels and external current on electrical behaviour in an excitable membrane....

%We investigate in detail the influence of sodium current conductance and potassium current conductance on the dynamical behaviour of the modified model. Variation of sodium current conductance changes the dynamics qualitatively. We perform a numerical bifurcation analysis of the model with sodium and potassium current conductances as bifurcation parameters. \textcolor{blue}{The bifurcation of solutions varying sodium current conductance produces complex bifurcation structure that is not present in the existing results of original Morris-Lecar model.}
 
% \PACS{05.45.Xt Synchronization, coupled oscillators \and 05.45.−a Nonlinear dynamics and chaos \and
% 47.20.Ky Nonlinearity, bifurcation, and symmetry breaking \and 89.75.Fb Structures and organization in
% complex systems}
\end{abstract}

\section{Introduction}
The variation in concentration of ions across the cell membrane results in fluxes of ions through the voltage-gated ion channels. This electrophysiological process in the cell membrane plays a fundamental role in understanding the electrical activities in excitable cells such as neurons \citep{Mondal2019BifurcationModel}, muscle cells \citep{Gonzalez-Fernandez1994} and hormones \citep{Karl}. The temporal variation of the cell membrane potential due to external stimulation is known as an action potential. Different ion channels play different roles in the generation of an action potential. Depending on the cell, the opening of ${\rm Na}^{+}$ (${\rm Ca}^{2+}$) channels causes influx of ${\rm Na}^{+}$ (${\rm Ca}^{2+}$) and the membrane potential becomes more positive, hence the membrane is depolarised. When the K\textsuperscript{+} channels are open, there is efflux of K\textsuperscript{+} which results in the repolarisation of the cell. Later, the membrane potential becomes more negative than the resting potential and the membrane is hyperpolarised. At this stage, the membrane will not respond to stimulus until it returns to the resting potential \citep{Izhikevich2007DynamicalBursting,Ermentrout2008FoundationsNeuroscience,  JamesKeener2009,mythesis}.

 %Intensive physiological experiments have been carried out in investigate the underlying mechanisms of interactions between ion channels and action potentials.
 From the viewpoint of mathematics, numerous mathematical models have been developed to study the nonlinear dynamics involved in the generation of an action potential in the cell membrane. They are often modelled by a nonlinear system of ordinary differential equations (ODEs). Among the famous works is the one by Hodgkin and Huxley (\citeyear{Hodgkin1952}) on the conduction of electrical impulses along a squid giant axon. In their experiments, it was reported that action potentials depends on the influx of ${\rm Na}^{+}$. This work laid foundation for other electrophysiological models. Other well-known models are the FitzHugh-Nagumo model (\citeyear{Fitzhugh, Nagumo1962}), the Morris-Lecar (ML) model (\citeyear{Morris}), the Chay model (\citeyear{Chay1985ChaosCell}), and the Smolen and Keizer model (\citeyear{smolen}).  

ML model describes the electrical activities of a giant barnacle muscle fibre membrane. Despite being a model for muscle cell, it has been widely used in modelling electrical activities in other excitable cells mostly in neurons \citep{azizi, Jia2018, Prescott2, Zhao2017TransitionsAutapse}. Based on experimental observations, ML model is formulated on the assumption that the electrical activities in barnacle muscle depend largely on fluxes of ${\rm Ca}^{2+}$ and  ${\rm K}^{+}$ rather than ${\rm Na}^{+}$. On this basis, their model consists of three ODEs. It is observed that the ${\rm Ca}^{2+}$ current activates faster than the ${\rm K}^{+}$ current and the charging capacitor \citep{keynes73}. Thus, the model is further reduced to two ODEs by setting the ${\rm Ca}^{2+}$ activation to quasi-steady state.

The two-dimensional ML model has been extensively used in many single-cell models \citep{Hetongwang2011, Lv2016, Upadhyay2017,hammed} and network of cells \citep{Fujii,Lafranceschina2014,Meier,Hartle17} studies despite it is an approximation of the three-dimensional ML model. In spite of little attention to the three-dimensional model, it has been used in modelling electrophysiological studies. For example, Gottschalk and Haney (\citeyear{Gottschalk}) investigated how the activity of the ion channels are regulated by anaesthetics. The three-dimensional ML model was used by Marreiros et al (\citeyear{Marreiros}) for modelling dynamics in neuronal populations using a statistical approach. Also, Gonz\'{a}lez-Miranda (\citeyear{Gonzalez-Miranda2014}) investigated pacemaker dynamics in ML model using the three-dimensional model. Gall and Zhou (\citeyear{Gall}) considered four-dimensional ML model by including the second inward ${\rm Na}^{+}$ current.

Many recent papers have studied modified ML model by adding relevant inward and outward ionic currents \citep{Prescott2008BiophysicalInitiation,Duan2010,Meier,Bao2019,azizi2}. \cite{Zeldenrust2013} extended the ML model by including three additional ionic membrane currents: a T-type calcium current, a cation selective h-current and a calcium dependent potassium current to investigate reliability of spikes in thalamocortical relay cells. Also, \cite{azizi} added calcium dependent potassium current to the ML model to study bursting properties in neurons. They showed that the model has  complex dynamical behaviour including square-wave, elliptic, and parabolic busters depending on parameter combinations. \cite{Rajagopal2021} modified the ML model by incoporating the influence of electric and magnetic field on dynamical behaviours of network of neurons. They found complex spatiotemporal dynamics including chaotic bursting and spiral waves.

The purpose of this paper is to investigate the influence of sodium inward currents on variation of membrane voltage of a single excitable cell. In recent years, experimental and computational analyses have suggested that sodium currents are relevant in the generation of action potential in some muscle cells \citep{Nagata2004,Berra-Romani2005TTX-sensitiveCells,Ulyanova2018Voltage-dependentArterioles}. Bifurcation analysis is often used to investigate the mode of transition of electrical activities of excitable cells. It helps us to identify the key parameters that cause changes in the dynamical behaviour  qualitatively \citep{KuznetsovY.A.1995ElementsTheory, JamesKeener2009}. A lot of studies on bifurcation analyses have been carried out on the two-dimensional \citep{Govaerts2005TheApproach,Tsumoto2006BifurcationsModel,Prescott2008BiophysicalInitiation,hammed} and three-dimensional \citep{Gonzalez-Miranda2014} ML models, however, to our knowledge there appears no work in the literature that has extensively considered the bifurcation analysis of the four-dimensional ML model. In this present paper we focus on the maximum conductances of ion currents and external current as bifurcation parameters. As a consequence, we show some additional bifurcation that are not present in the existing results of ML model.

%The results of this paper are presented 
The paper is organised as follows. In Sect.~\ref{sec:model}, we present the model equations and the dynamics of the model upon variation of model parameters. A detailed bifurcation analyses are carried out
in Sect.~\ref{sec:bifurcation}. Finally, the conclusion is presented in Sect.~\ref{sec:conclusion}.

\section{Model Equation}\label{sec:model}
The classical Morris-Lecar (ML) model (\citeyear{Morris}) is a three-dimensional nonlinear system of ODEs, which is described as
\begin{align}
 \label{eq:1.1}
    \textrm{C}\frac{dV}{dt}&=I_{\rm ext}-I_{\rm L}-I_{\rm Ca}-I_{\rm K},\\
    \label{eq:2.1}
     \frac{dm}{dt}&=\lambda_m(V)(m_{\infty}(V)-m),\\
     \label{eq:3.1}
     \frac{dn}{dt}&=\lambda_n(V)(n_{\infty}(V)-n),
 \end{align}
 where $V$ is the membrane potential, $I_{\rm ext}$ is the external current,  and ${\rm C}$ is the membrane capacitance. $m$ and $n$ are the fraction of open calcium and potassium channels, respectively. The ionic currents in \eqref{eq:1.1} are defined as
\begin{equation}
\begin{aligned}
I_{\rm L}=g_{\rm L}(V-v_{\rm L}),~I_{\rm Ca}=g_{\rm Ca}m(V-v_{\rm Ca}),~I_{\rm K}=g_{\rm K}n(V-v_{\rm K}),
\end{aligned}
\end{equation}
where $g_{\rm L}$, $g_{\rm Ca}$, and $g_{\rm K}$ are the maximum conductances of the leak, calcium, and potassium channels, respectively. Also $v_{\rm L}$, $v_{\rm Ca}$, and $v_{\rm K}$ are the Nerst reversal potentials of the leak, ${\rm Ca}^{2+}$, ${\rm K}^{+}$ and ${\rm Na}^{+}$  channels, respectively. 

\begin{figure}[htbp]
    \centering
    \includegraphics[scale=0.6]{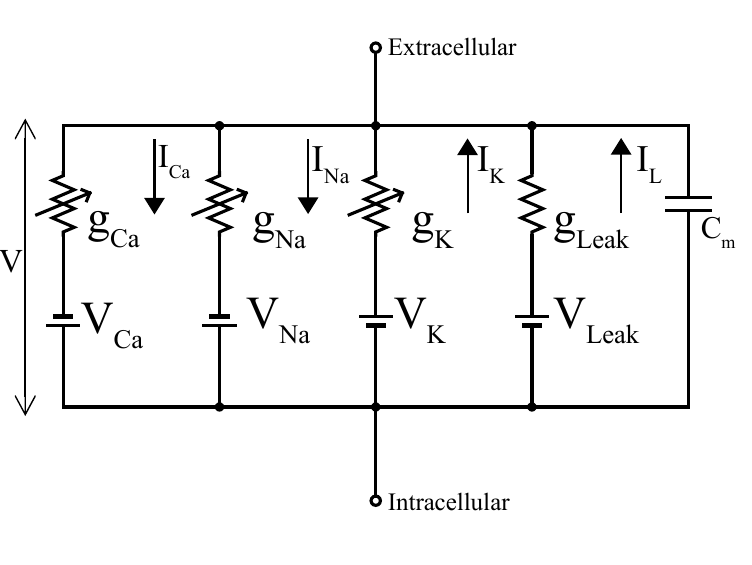}
    \caption{Equivalent circuit representation of the cell membrane with four ionic channels}
    \label{fig:circuit}
\end{figure}

Taking into account the contribution of ${\rm Na}^{+}$ on membrane depolarisation, we extend the ML model by adding ${\rm Na}^{+}$ current, $I_{\rm Na}=g_{\rm Na}w(V-v_{\rm Na})$, in \eqref{eq:1.1}. With this current the ML model becomes a four-dimensional system of ODEs defined as
 \begin{align}
 \label{eq:1.1a}
    \textrm{C}\frac{dV}{dt}&=I_{\rm ext}-I_{\rm L}-I_{\rm Ca}-I_{\rm K}-I_{\rm Na},\\
    \label{eq:2.1a}
     \frac{dm}{dt}&=\lambda_m(V)(m_{\infty}(V)-m),\\
     \label{eq:3.1a}
     \frac{dn}{dt}&=\lambda_n(V)(n_{\infty}(V)-n),\\
     \label{eq:4.1a}
    \frac{dw}{dt}&=\lambda_w(V)(w_{\infty}(V)-w).
 \end{align}
 The equivalent circuit representation of the cell membrane with four ionic channels, $I_{\rm L}$, $I_{\rm Ca}$, $I_{\rm K}$, and $I_{\rm Na}$, is shown in Fig.~\ref{fig:circuit}.
The fraction of open ${\rm Ca}^{2+}$, ${\rm K}^{+}$ and ${\rm Na}^{+}$ channels at steady state, denoted by $m_{\infty}$, $n_\infty$, and $w_\infty$ are defined as
 \begin{equation*}
\begin{aligned}
 m_\infty(V)&=0.5\left(1+\tanh\left(\frac{V-\bar{v}_1}{\bar{v}_{2}}\right)\right),\\
n_{\infty}(V)&=0.5\left(1+\tanh\left(\frac{V-\bar{v}_3}{\bar{v}_{4}}\right)\right),\\
w_{\infty}(V)&=0.5\left(1+\tanh\left(\frac{V-\bar{v}_5}{\bar{v}_{6}}\right)\right).
\end{aligned}
\end{equation*}
The voltage-dependent rate constants associated with calcium, potassium and sodium channels are 
\begin{equation*}
\begin{aligned}
\lambda_{m}(V)&=\psi_{m}\cosh\left(\frac{V-\bar{v}_{1}}{2\bar{v}_{2}}\right),\\
\lambda_{n}(V)&=\psi_{n}\cosh\left(\frac{V-\bar{v}_{3}}{2\bar{v}_{4}}\right),\\
\lambda_{w}(V)&=\psi_{w}\cosh\left(\frac{V-\bar{v}_{5}}{2\bar{v}_{6}}\right),
\end{aligned}
\end{equation*}
Unless stated otherwise, parameter values are as listed in \cite{Gall}: $\textrm{C}=1$, $I_{\text{ext}}=50$, $\textrm{g}_{\text{L}}=2$, $v_{\textrm{L}}=-50$, $\textrm{g}_{\textrm{Ca}}=4$, $v_{\text{Ca}}=100$, $\mathrm{g_K}=8$, $v_{K}=-70$, $\textrm{g}_{\textrm{Na}}=2$, $v_{\textrm{Na}}=55$, $v_{1}=-1$, $v_{2}=15$, $v_{3}=10$, $v_4=14.5$, $v_5=5$, $v_6=15$, $\psi_{m}=1$, $\psi_{n}=0.0667$, $\psi_{w}=0.033$.

\subsection{\textbf{Changes to Excitable Dynamics as a Parameter is Varied}}\label{sec:dynamics}
\begin{figure}[htbp]
\centering
  \begin{subfigure}[b]{.5\linewidth}
    \centering
    \caption{}
    \includegraphics[width=\textwidth]{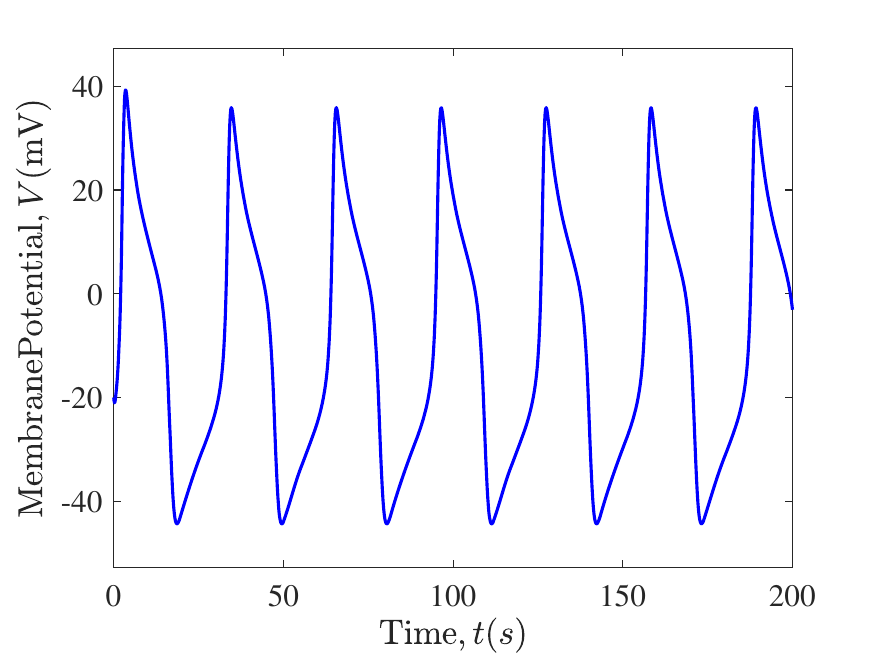}
    \label{fig:gca_absent}
  \end{subfigure}%  
  \begin{subfigure}[b]{.5\linewidth}
    \centering
    \caption{}
    \includegraphics[width=\textwidth]{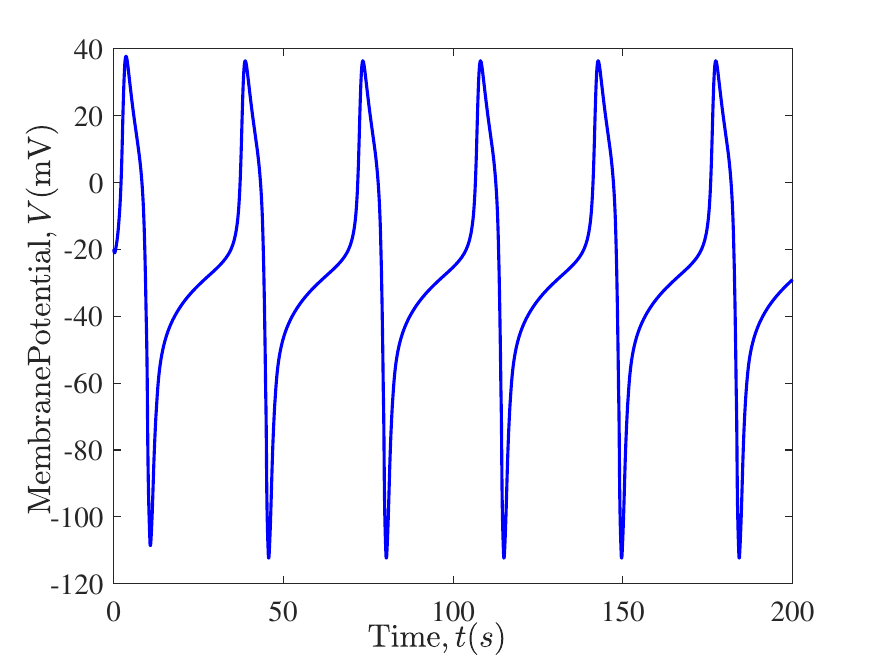}
    \label{fig:gca_present}
  \end{subfigure}%  
  \label{fig:gna-blocked}
  \caption{Time series of the membrane potential $V$ when the ${\rm Na}^{+}$ conductance ${\rm g}_{Na}$ is: (a) blocked ($g_{\rm Na}=0$); (b) unblocked ($g_{\rm Na}\ne0$)}
  \end{figure}

%of sodium $({\rm Na}^{+})$ current on the membrane potential by varying its conductance $g_{\rm Na}$.
To analyse the model, we first assess the effects of ${\rm Na}^{+}$ current on electrical activity.  To do this, we block the conductance $g_{\rm Na}$ for the ${\rm Na}^{+}$ current. The model is integrated numerically using the standard fourth-order Runge–Kutta method using a step size of 0.05 in the numerical software XPPAUT \citep{Ermentrout2002SimulatingStudents}. Fig.~\ref{fig:gca_absent} and \ref{fig:gca_present} show the time series of the membrane potential $V$ for model \eqref{eq:1.1a}--\eqref{eq:4.1a} when the ${\rm Na}^{+}$ conductance is blocked and unblocked, respectively. Over a range of parameters considered, we found that the addition of ${\rm Na}^{+}$ current causes the membrane potential to shift to more hyperpolarised values for hyperpolarised states, see Fig.~\ref{fig:gca_present}. This tells us that the effects of ${\rm Na}^{+}$ conductance is non-negligible.

\begin{figure}[htbp]
\centering
  \begin{subfigure}[b]{.4\linewidth}
    \centering
    \caption{}
   \includegraphics[width=\textwidth]{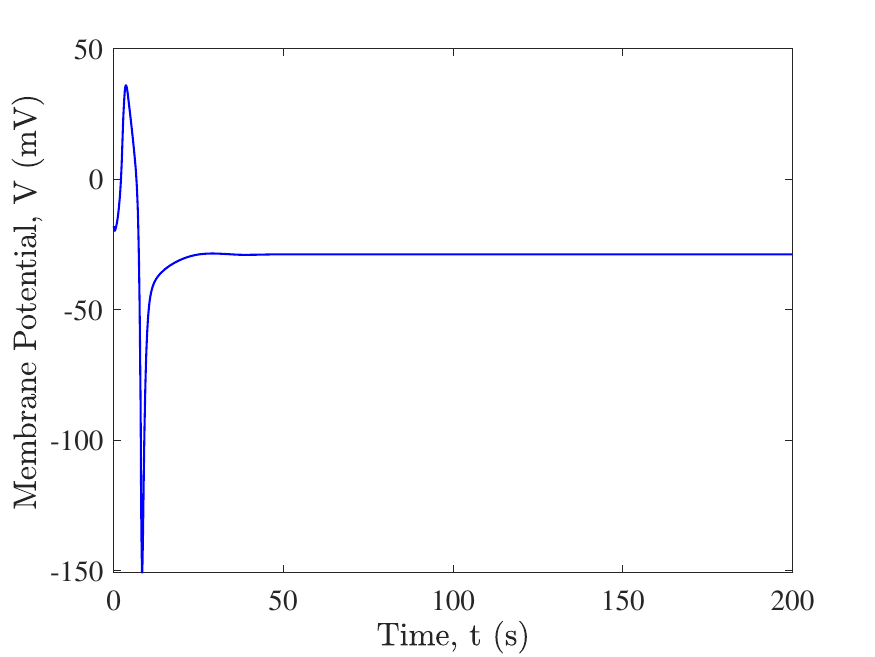}
    \label{fig:TTgna20}
  \end{subfigure}%  
  \begin{subfigure}[b]{.4\linewidth}
    \centering
    \caption{}
    \includegraphics[width=\textwidth]{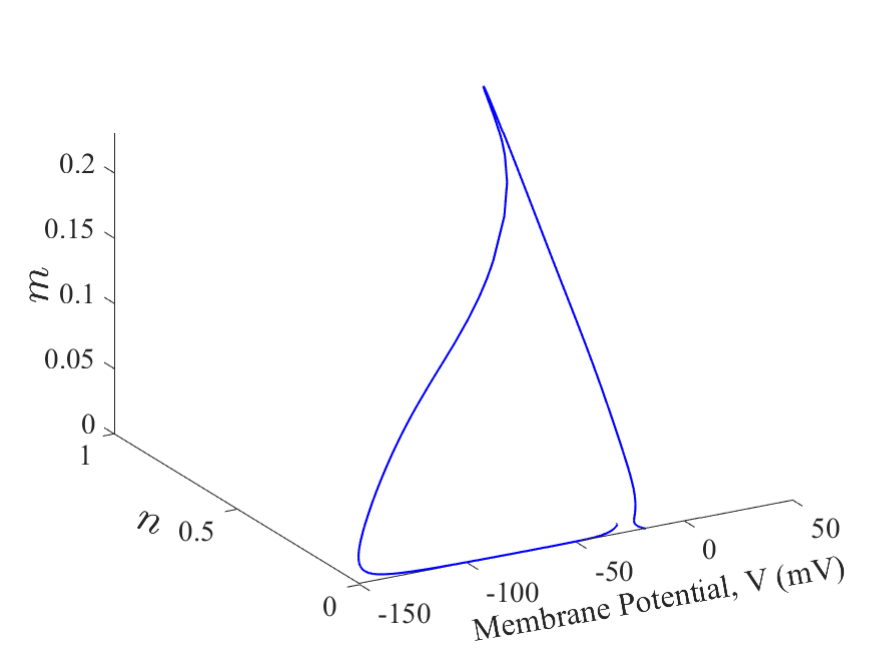}
    \label{fig:PPgna20}
  \end{subfigure}\\%  
  \begin{subfigure}[b]{.4\linewidth}
    \centering
    \caption{}
    \includegraphics[width=\textwidth]{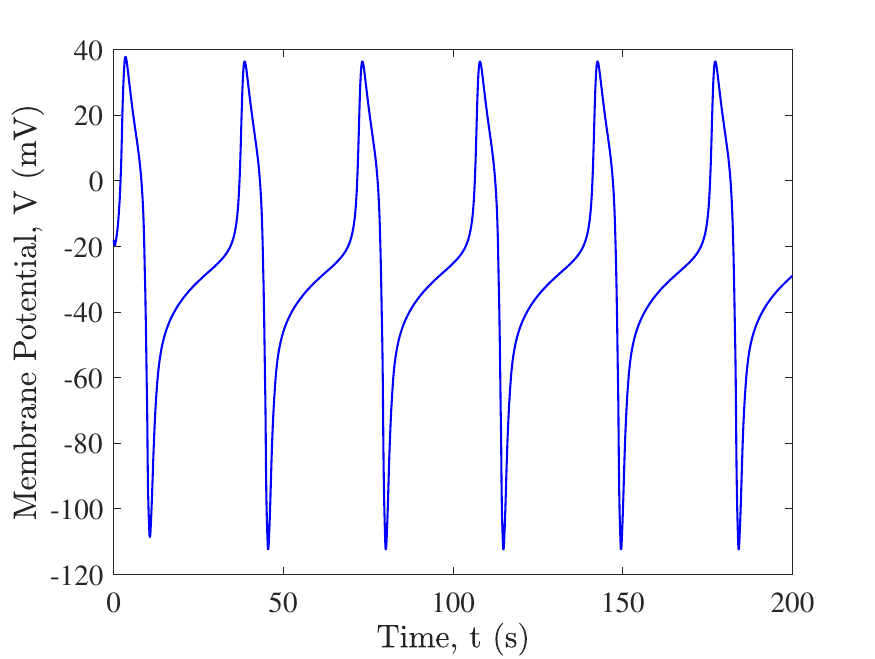}
    \label{fig:TTgna10}
  \end{subfigure}%
  \begin{subfigure}[b]{.4\linewidth}
    \centering
    \caption{}
    \includegraphics[width=\textwidth]{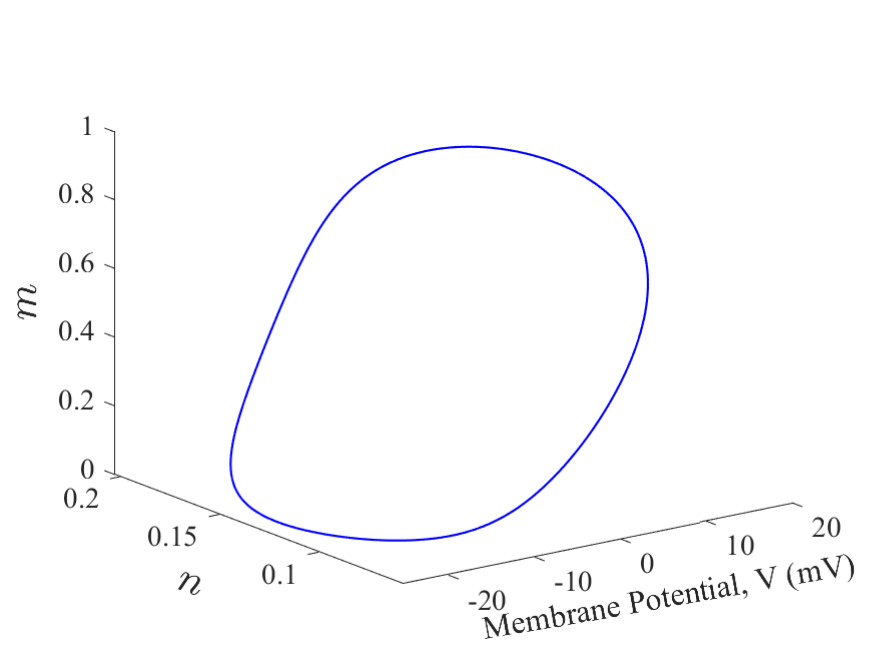}
    \label{fig:SLCgna}
  \end{subfigure}\\%
  \begin{subfigure}[b]{.4\linewidth}
    \centering
   \caption{}
   \includegraphics[width=\textwidth]{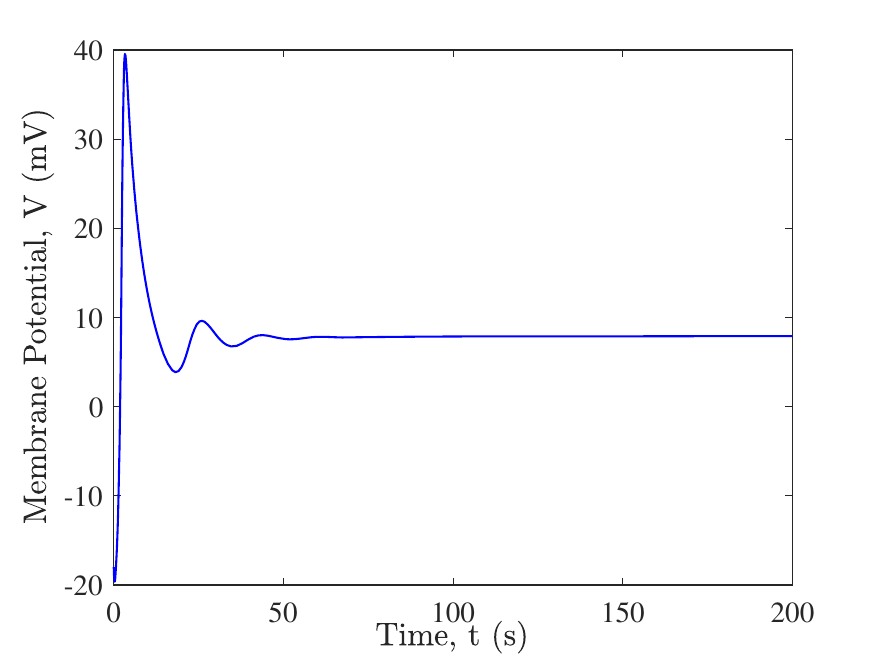}
   \label{fig:TTgna18}
  \end{subfigure}%
  \begin{subfigure}[b]{.4\linewidth}
    \centering
   \caption{}
   \includegraphics[width=\textwidth]{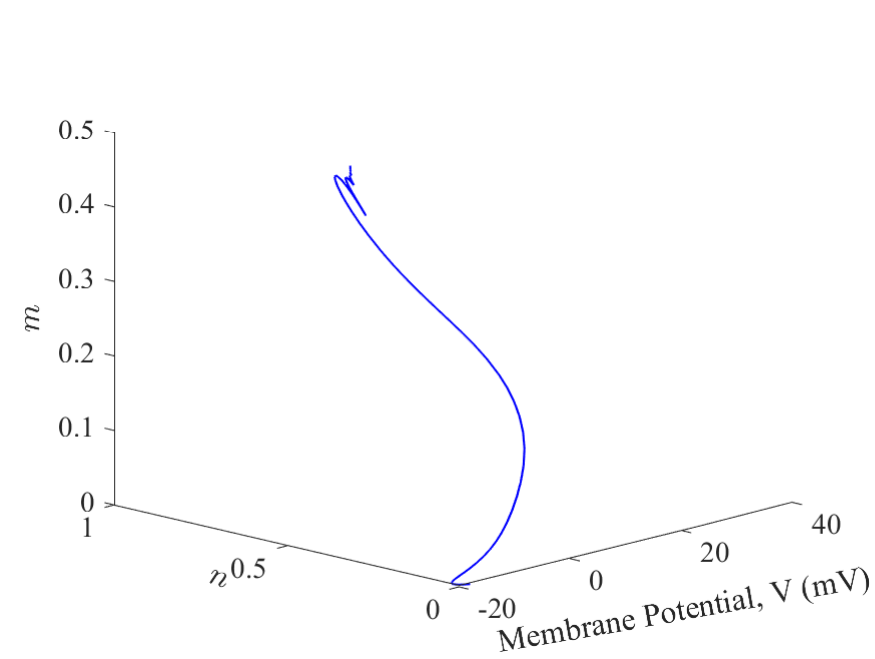}
   \label{fig:PPgna18}
  \end{subfigure}%
 \caption{Numerical simulations of the membrane potential $V$ for (a) $g_{\rm Na}=-20$; (c) $g_{\rm Na}=-10$; (e) $g_{\rm Na}=1.8$. Their corresponding phase space are (b), (d) and (f), respectively}
 \label{fig:TT_PPgna}
\end{figure}

As seen in previous studies \citep{Gonzalez-Miranda2014,hammed}, variation of parameters can result in changes to dynamical behaviour of the model, for example, transitions from rest state to periodic oscillations and vice versa. Here, we investigate the effects of maximum conductance on the dynamical behaviour of model \eqref{eq:1.1a}--\eqref{eq:4.1a}. The dynamics of the membrane potential $V$ upon varying ${\rm Na}^{+}$ current conductance $g_{\rm Na}$ is shown in Fig.~\ref{fig:TT_PPgna}. For the range of values of $g_{\rm Na}$ considered, the system either converge to a rest state or oscillatory state. For extremely low values of $g_{\rm Na}$, a single action potential is observed. In particular, the time evolution and its corresponding phase space for $g_{\rm Na}=-20$ are shown in  Figs.~\ref{fig:TTgna20} and \ref{fig:PPgna20}, respectively. Upon increasing $g_{\rm Na}$, periodic oscillations of action potentials are observed in the system, see Fig.~\ref{fig:TTgna10}. The periodic oscillations correspond to a closed loop in the phase space, see Fig.~\ref{fig:SLCgna}. The closed loop is also known as a limit cycle or periodic orbit. Further increasing $g_{\rm Na}$, the system stabilises to a steady state, see Figs.~\ref{fig:TTgna18} and \ref{fig:PPgna18}. Similar behaviours are observed when $g_{\rm K}$ and $g_{\rm Ca}$ are varied (results not shown).  A detailed bifurcation analysis is given in Sec.~\ref{sec:bifurcation} to further understand how the dynamical properties of model \eqref{eq:1.1a}--\eqref{eq:4.1a} change as parameter values is varied.

\section{Numerical Bifurcation Analysis}\label{sec:bifurcation}
With the aid of bifurcation analysis, we examine the dynamical behaviour of model \eqref{eq:1.1a}--\eqref{eq:4.1a} as different model parameters are varied in turn. The bifurcation diagrams are produced in XPPAUT and edited in MATLAB. The continuation parameters used in XPPAUT are $\verb|NTST=100|$, $\verb|NMAX=2000|$, $\verb| Method=stiff|$, $\verb|EPSL=1e-7|$ , $\verb|EPSU=1e-7|$, $\verb|EPSS=1e-7|$, $\verb|ITMX=20|$, $\verb|ITNW=20|$, $\verb|DSMIN=1e-05|$, $\verb|DSMAX=0.05|$. The abbreviations and labels for the bifurcation points are given in Table~\ref{Tab:bifpoints}.
\begin{table}[htbp]
\centering
\caption{Abbreviations and notations  of bifurcation points}
\label{Tab:bifpoints}       
\begin{tabular}{ll}
\hline\noalign{\smallskip}
Bifurcation & Abbreviation  \\
\noalign{\smallskip}\hline\noalign{\smallskip}
Hopf bifurcation & HB  \\[1mm]
Saddle-node bifurcation & SN   \\[1mm]
Saddle-node bifurcation of cycles & SNC \\[1mm]
Homoclinic bifurcation & HC  \\[1mm]
Period-doubling bifurcation & PD \\[1mm]
\noalign{\smallskip}\hline
\end{tabular}
\end{table}

\subsection{\textbf{Influence of $g_{\rm Na}$}}\label{sec:gnabif}
\begin{figure}[htbp]
\centering
  \begin{subfigure}[b]{.5\linewidth}
    \centering
    \caption{}
    \includegraphics[width=\textwidth]{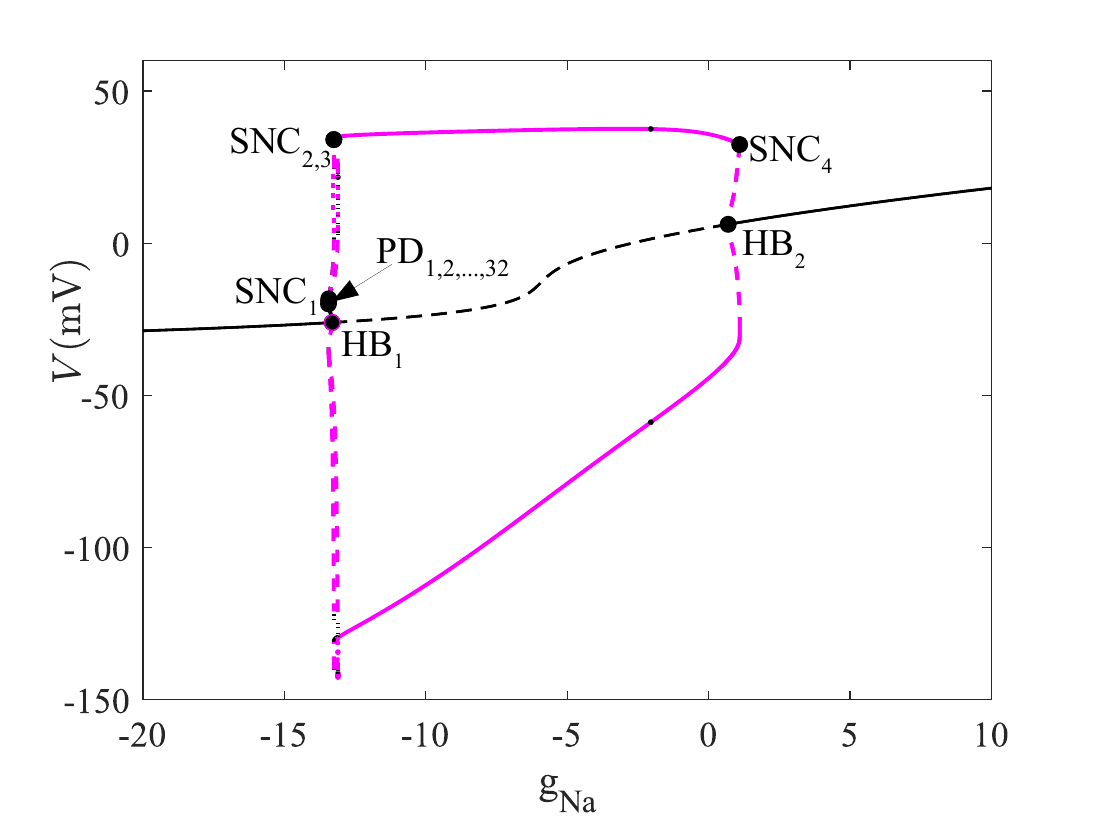}
    \label{fig:gnabif}
  \end{subfigure}\\%  
  \begin{subfigure}[b]{.5\linewidth}
    \centering
    \caption{}
    \includegraphics[width=\textwidth]{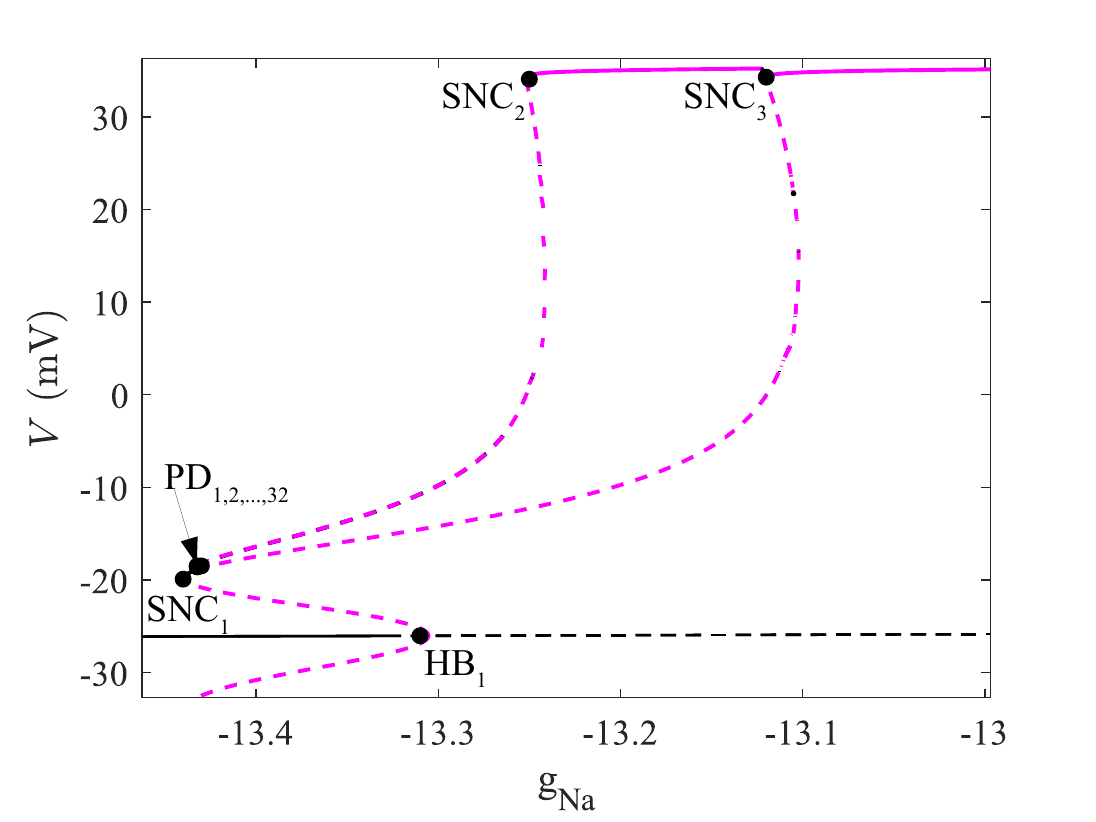}
    \label{fig:gnabifzoom}
  \end{subfigure}%  
  \begin{subfigure}[b]{.5\linewidth}
    \centering
    \caption{}
    \includegraphics[width=\textwidth]{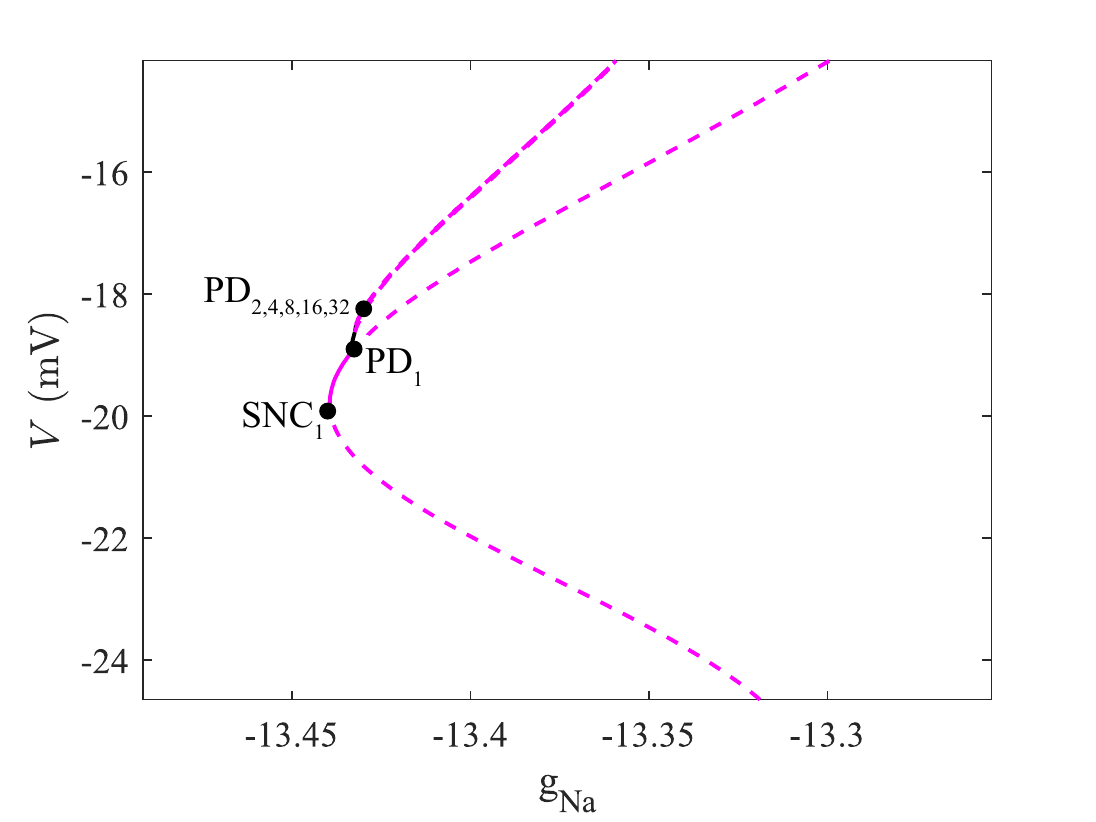}
    \label{fig:gnabifzoom1}
  \end{subfigure}% 
  \caption{(a) Bifurcation diagram of the membrane potential $\mathrm{V}$ with $g_{\rm Na}$ as bifurcation parameter. The remaining parameter values are fixed as in Sec.~\ref{sec:model}. (b)-(c) are enlargements of (a). Continuous [dashed] curves correspond to stable [unstable] solutions.  Black [mangenta] curves correspond to equilibria [periodic oscillations]. HB: Hopf bifurcation; SN: saddle-node bifurcation (of an equilibrium); SNC: saddle-node bifurcation of a periodic orbit; PD: period-doubling bifurcation}
\label{fig:gnabifs}
  \end{figure}

Here, we vary $g_{\rm Na}$ to explore the effects of ${\rm Na}^{+}$ current on the dynamical behaviour of model \eqref{eq:1.1a}--\eqref{eq:4.1a}. Fig.~\ref{fig:gnabifs} is a bifurcation diagram of the membrane potential $V$ upon varying $g_{\rm Na}$ with other parameters fixed. For the range of values of $g_{\rm Na}$ considered, there exists a unique equilibrium. The system has a stable equilibrium except between two Hopf bifurcations where the equilibrium is unstable. As seen in Fig.~\ref{fig:gnabif}, the system loses stability through a subcritical Hopf bifurcation ${\rm HB}_{1}$ at $g_{\rm Na}\approx-13.305$ and regains stability at another subcritical Hopf bifurcation ${\rm HB}_{2}$ at $g_{\rm Na}\approx 0.69436$. The unstable limit cycle generated at ${\rm HB}_{1}$ gain stability through a saddle-node bifurcation of cycle ${\rm SNC}_{1}$ at $g_{\rm Na}\approx-13.4394$, and loses stability at a period-doubling bifurcation ${\rm PD}_{1}$. The unstable limit cycle branch regains stability through another ${\rm SNC}_{3}$ at $g_{\rm Na}\approx -13.1223$. The stable double-period limit cycle branch emanated from the ${\rm PD}_{1}$ loses stability at another period doubling bifurcation ${\rm PD}_{2}$ at $g_{\rm Na}\approx-13.4323$, and it regains stability through a ${\rm SNC}_{2}$ at $g_{\rm Na}\approx -13.2516$ before converging to the first unstable limit cycle branch at $g_{\rm Na}\approx-13.1223$, see Fig.~\ref{fig:gnabifzoom}. Upon further increasing the value of $g_{\rm Na}$, the limit cycle loses stability in a ${\rm SNC}_{4}$ at $g_{\rm Na}\approx 1.10527$ before it ends in a ${\rm HB}$ point at $g_{\rm Na}\approx 0.69436$. 

\begin{figure}[htbp]
\centering
  \begin{subfigure}[b]{.4\linewidth}
    \centering
    \caption{}
    \includegraphics[width=\textwidth]{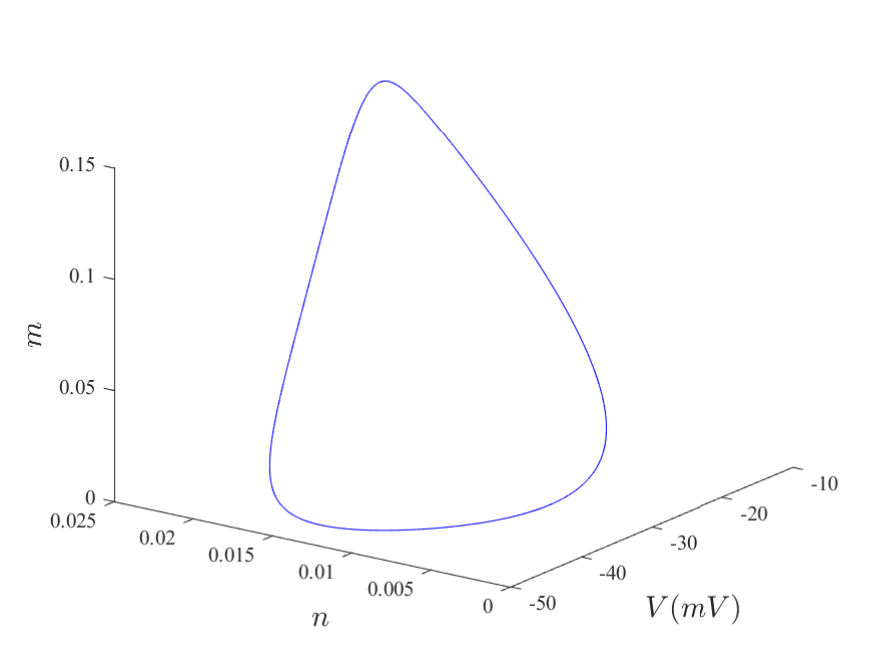}
    \label{fig:PD-1_gna}
  \end{subfigure}%  
  \begin{subfigure}[b]{.4\linewidth}
    \centering
    \caption{}
    \includegraphics[width=\textwidth]{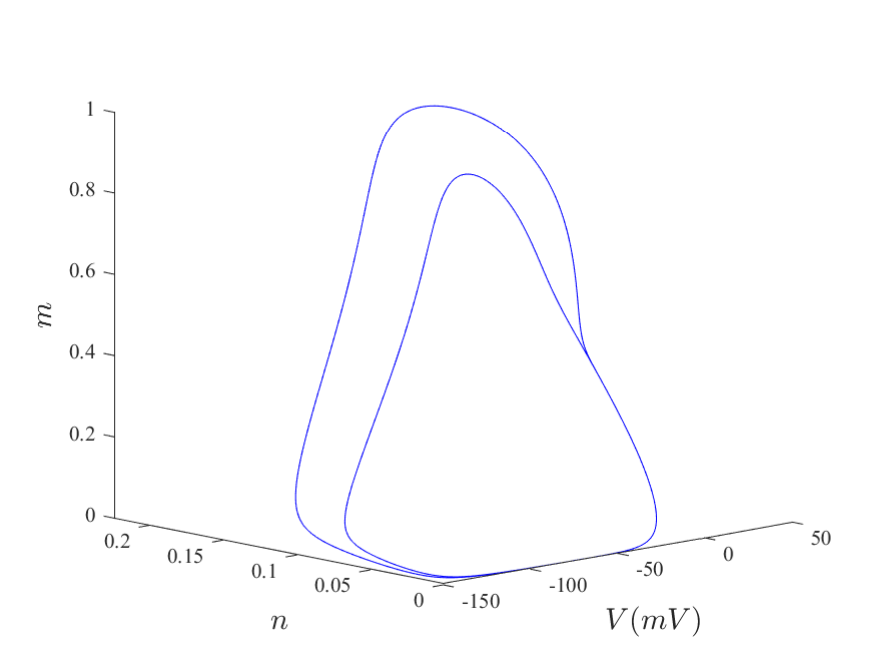}
    \label{fig:PD-2_gna}
  \end{subfigure}\\%  
  \begin{subfigure}[b]{.4\linewidth}
    \centering
    \caption{}
    \includegraphics[width=\textwidth]{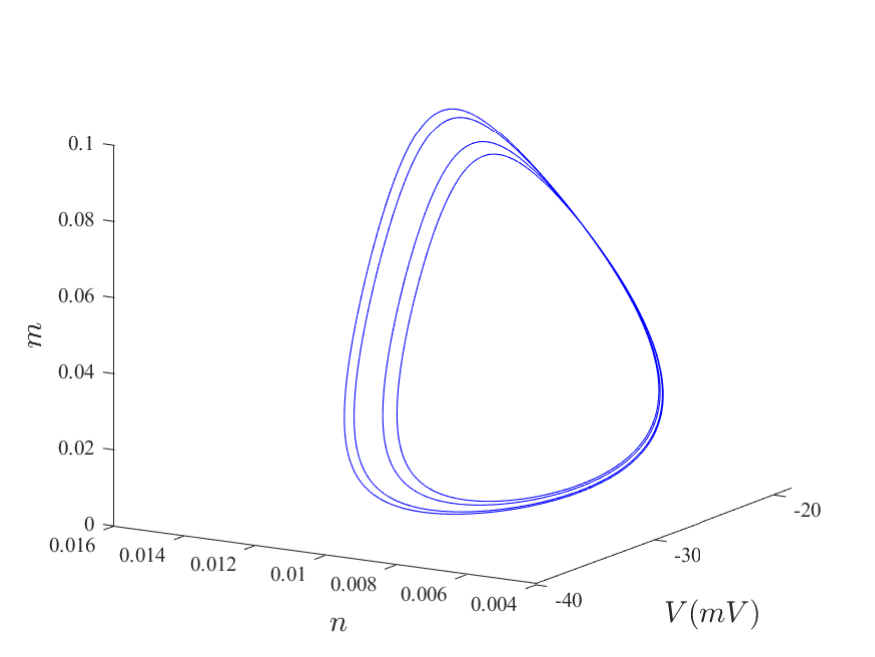}
    \label{fig:PD-4_gna}
  \end{subfigure}%
  \begin{subfigure}[b]{.4\linewidth}
    \centering
    \caption{}
    \includegraphics[width=\textwidth]{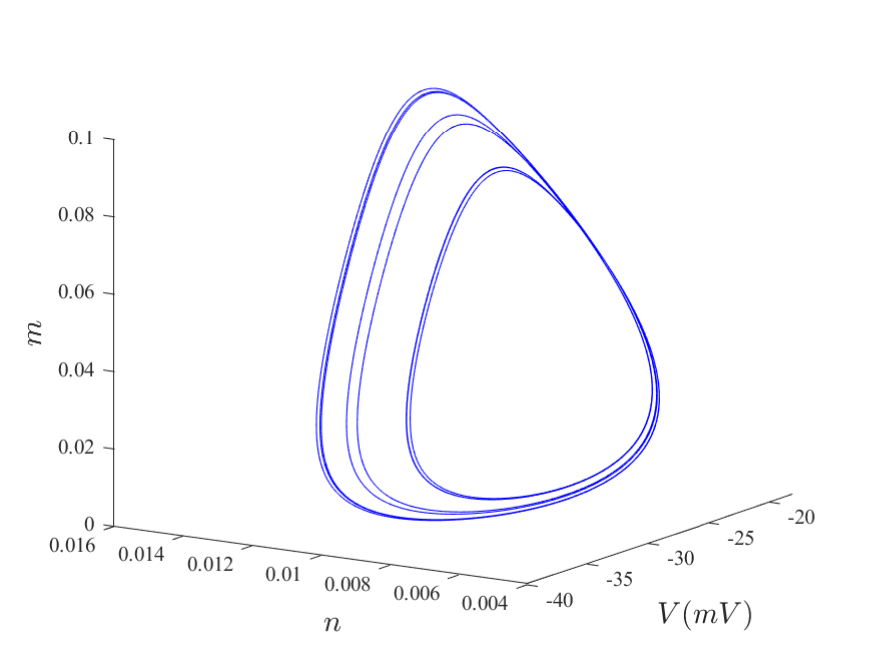}
    \label{fig:PD-8_gna}
  \end{subfigure}\\%
  \begin{subfigure}[b]{.4\linewidth}
    \centering
   \caption{}
   \includegraphics[width=\textwidth]{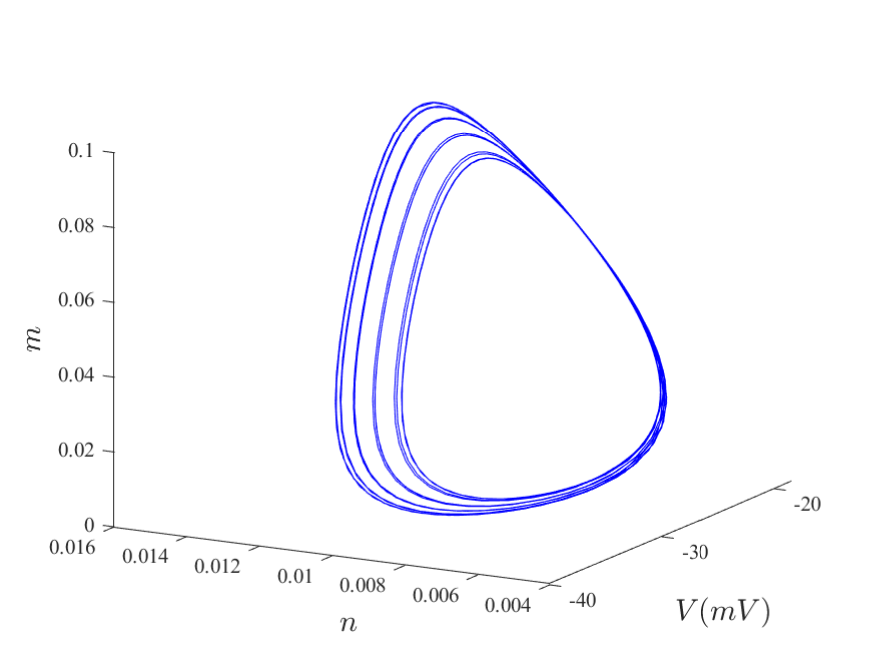}
   \label{fig:PD-16_gna}
  \end{subfigure}%
  \begin{subfigure}[b]{.4\linewidth}
    \centering
   \caption{}
   \includegraphics[width=\textwidth]{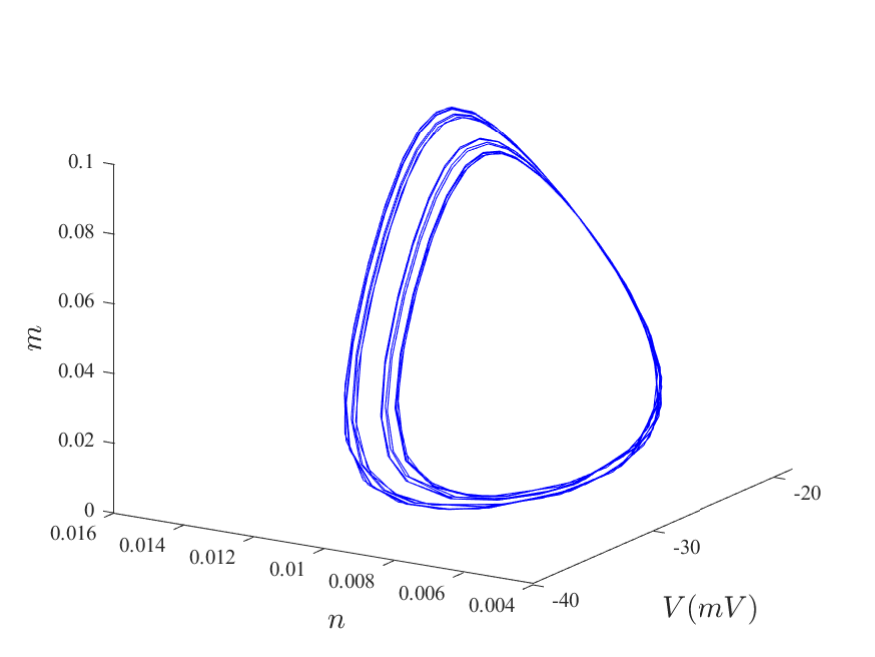}
   \label{fig:PD-32_gna}
  \end{subfigure}%
 \caption{Phase-space of \eqref{eq:1.1a}--\eqref{eq:4.1a} showing the cascade of period-doubling bifurcations. (a) Period-1 (b) Period-2 (c) Period-4 (d) Period-8 (e) Period-16 (f) Period-32, respectively}
 \label{fig:perioddoublinggna}
\end{figure}

Continuation of ${\rm PD}_{2}$ bifurcation results in another stable limit cycle that loses stability at a period doubling bifurcation ${\rm PD}_{4}$, the period of this limit cycle is double the period of the limit cycle of ${\rm PD}_{2}$. Continuing this process results in a cascade of PD bifurcations of limit cycles, and this may lead to chaotic dynamics in the system \citep{Seydel2010PracticalAnalysis,Kugler2017PeriodAfterdepolarizations}. Table~\ref{Tab:PD} shows the values and period of the period doubling bifurcations that arise as $g_{\rm Na}$ is varied. The projection of the periodic trajectories for Period-1, 2, 4, 8, 16 and 32 onto $(V,n,m)$ phase space is illustrated in Fig.~\ref{fig:perioddoublinggna}. All the double-period unstable limit cycles generated at each PD points undergo SNC bifurcations before they converge to the limit cycle emanated from the first ${\rm HB}$ bifurcation. 
  \begin{table}[htbp]
\centering\caption{Summary of the parameter values and period of Period doubling bifurcations that arise as $g_{\rm Na}$ is varied}
\label{Tab:PD}       
\begin{tabular}{lll}
\hline\noalign{\smallskip}
Bifurcation & \hspace{3mm} $g_{\rm Na}$ & Period \\
\noalign{\smallskip}\hline\noalign{\smallskip}
${\rm PD}_{1}$ & -13.4334&  36.0272\\[1mm]

${\rm PD}_{2}$ & -13.4323 &  72.1846\\[1mm]

${\rm PD}_{4}$ & -13.4321& 144.489 \\[1mm]

${\rm PD}_{8}$&  -13.4320&  289.001\\[1mm]

${\rm PD}_{16}$ & -13.4320 &  578.025\\[1mm]

${\rm PD}_{32}$ & -13.4320 &  1156.05 \\[1mm]
\noalign{\smallskip}\hline
\end{tabular}
\end{table}
  
\subsection{\textbf{Influence of $g_{\rm K}$ and $g_{\rm Ca}$}}\label{sec:gkbif}
\begin{figure}[htbp]
\centering
  \begin{subfigure}[b]{.5\linewidth}
    \centering
    \caption{}
    \includegraphics[width=\textwidth]{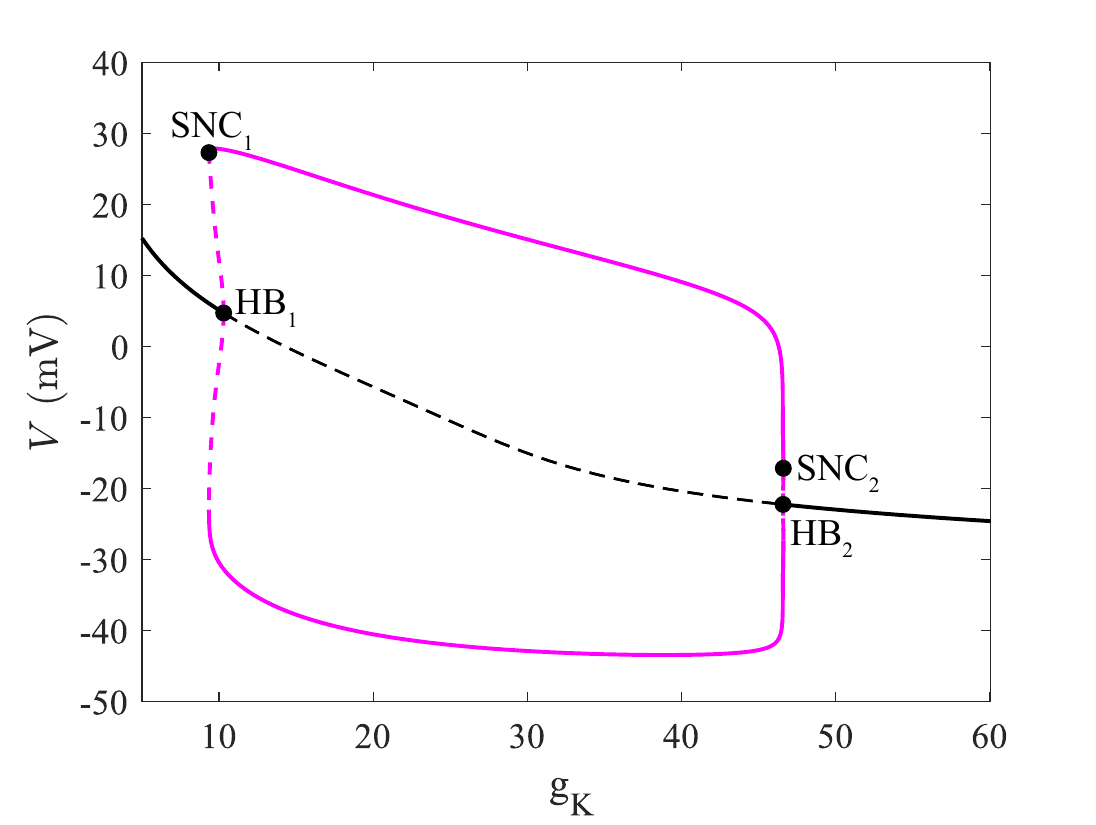}
    \label{fig:gKbif}
  \end{subfigure}%  
  \begin{subfigure}[b]{.5\linewidth}
    \centering
    \caption{}
    \includegraphics[width=\textwidth]{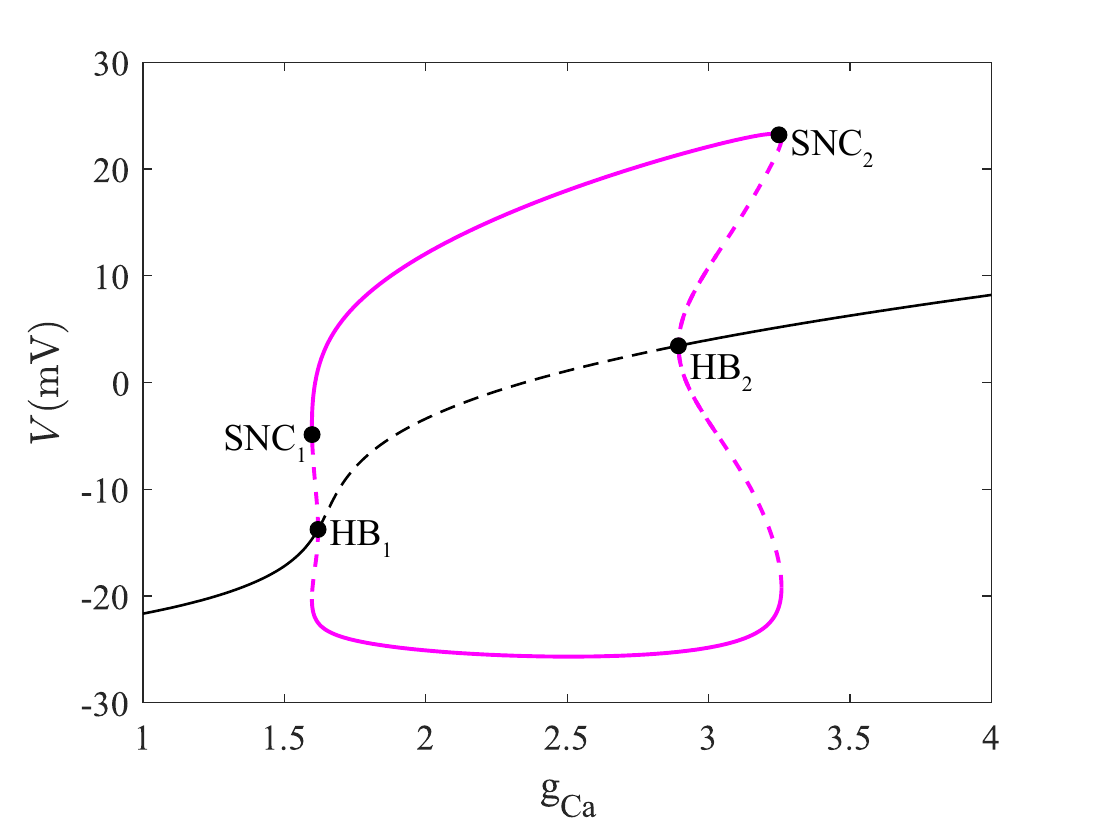}
    \label{fig:gca_bif}
  \end{subfigure}%  
  \caption{Bifurcation diagrams of the membrane potential $V$ with (a) $\mathrm{g_{K}}$ (b) $g_{\text{Ca}}$ as the bifurcation parameters and other parameters are fixed as in Sec.~\ref{sec:model}. The labels and other conventions are as in Fig.~\ref{fig:gnabifs}}
  \end{figure}

 Fig.~\ref{fig:gKbif} shows the bifurcation diagram of the membrane potential $V$ as $g_{\rm K}$ is varied. For the values of $g_{\rm K}$ considered, there exists a unique equilibrium. For extremely low values and high values of $g_{\rm K}$, the equilibrium is stable. Increasing $g_{\rm K}$, the system loses stability through a subcritical Hopf bifurcation ${\rm HB}_{1}$ at $g_{\rm K}\approx10.029$ and this leads to emergence of an unstable limit cycle which becomes stable through a saddle node bifurcation of cycles ${\rm SNC}_{1}$ at $g_{\rm K}\approx9.345$. As $g_{\rm K}$ increases further, the stable limit cycle changes stability in another saddle node bifurcation of cycles ${\rm SNC}_{2}$ at $g_{\rm K}\approx46.598$. The unstable limit cycle ends in another subcritical Hopf bifurcation ${\rm HB}_{2}$ at $g_{\rm K}\approx 42.583$. Bistability is observed, that is, a stable limit cycle coexists with a stable equilibrium when $9.345\leq g_{\rm K}\leq 10.029$ and  $42.583\leq g_{\rm K} \leq 46.598$. %For these values of $g_{\rm K}$, . %The onset of oscillation is at a nonzero frequency, this is typical of oscillations that occur through a Hopf bifurcation and this type of behaviour is classified as Type II excitability by \cite{Rinzel1999AnalysisNetwork}.
  
Next, we vary the value of the parameter $g_{\rm Ca}$. Fig.~\ref{fig:gca_bif} shows the bifurcation diagram of the membrane potential $V$ as $g_{\rm Ca}$ is varied. As $g_{\rm Ca}$ is varied, the system loses stability through a subcritical Hopf bifurcation ${\rm HB}_{1}$ at $g_{\rm Ca}\approx1.6191$ and this results in emergence of unstable limit cycle which becomes stable through a saddle node bifurcation of cycles ${\rm SNC}_{1}$ at $g_{\rm Ca}\approx1.5974$. As $g_{\rm Ca}$ increases further, the stable limit cycle loses stability in another saddle-node bifurcation ${\rm SNC}_{2}$ at $g_{\rm Ca}\approx3.2579$ and the unstable limit cycle ends in a subcritical Hopf bifurcation ${\rm HB}_{2}$ at $g_{\rm Ca}\approx2.8938$. Between the two subcritical Hopf bifurcations, there exists a unique unstable equilibrium point. For $1.5974\leq g_{\rm Ca}\leq 1.6191$ and  $2.8938\leq g_{\rm Ca} \leq 3.2579$, a stable limit cycle coexists with a stable equilibrium and the system is bistable. For these values of $g_{\rm Ca}$, a stable limit cycle coexists with a stable equilibrium.

 \subsection{\textbf{Influence of $I_{\rm ext}$}}\label{sec:Iext}
\begin{figure}[htbp]
\centering
  \begin{subfigure}[b]{.5\linewidth}
    \centering
    \caption{}
    \includegraphics[width=\textwidth]{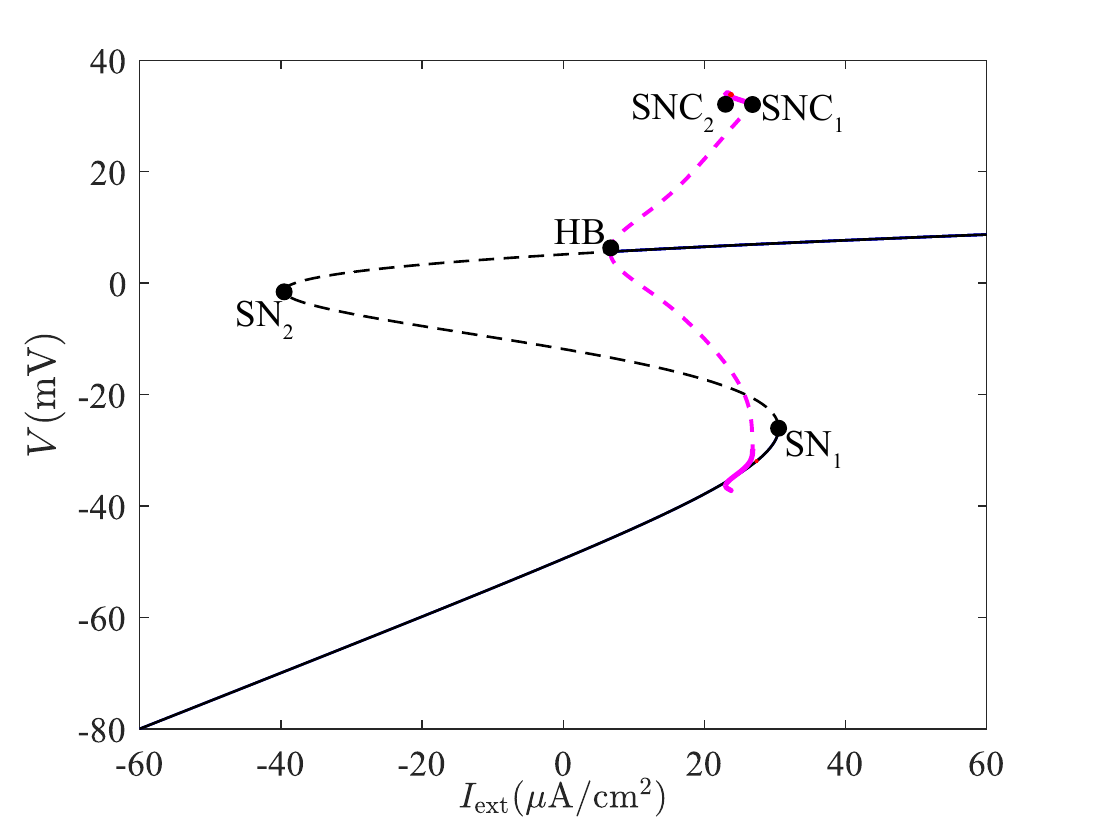}
    \label{fig:Ibif15}
  \end{subfigure}%  
  \begin{subfigure}[b]{.5\linewidth}
    \centering
    \caption{}
   \includegraphics[width=\textwidth]{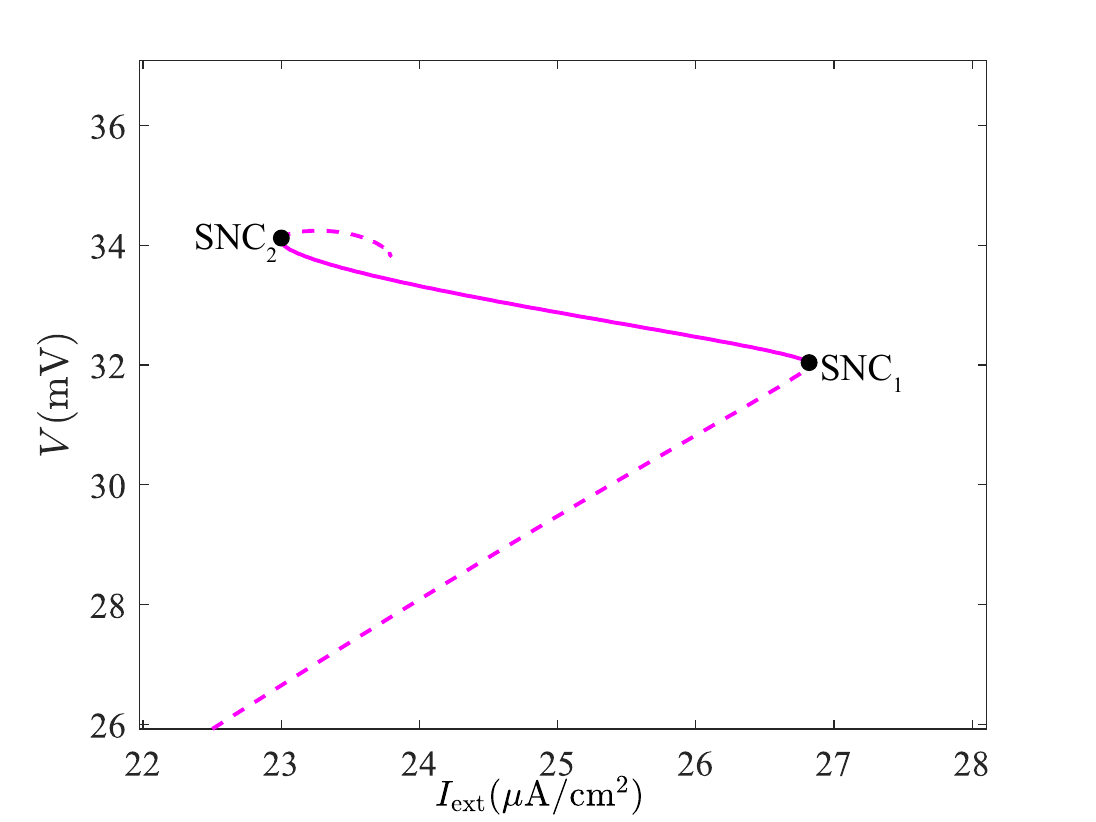}
    \label{fig:zoomIbif15}
  \end{subfigure}%  
  \caption{(a) Bifurcation diagram of the membrane potential $V$ with $\text{I}_{\text{ext}}$ as the bifurcation parameter. and other parameters are fixed as in Sec.~\ref{sec:model}. The labels and other conventions are as in Fig.~\ref{fig:gnabifs} }
  \end{figure}

Apart from maximum conductance of ionic channels, the influence of external current is highly important while investigating the dynamics of action potentials in electrophysiological studies. Here, we consider the effects of $I_{\rm ext}$ using two parameter sets. For set I, the parameter values are as listed in Sect.~\ref{sec:model}. Fig.~\ref{fig:Ibif15} is a bifurcation diagram of the membrane potential $V$ with the applied current $I_{\rm ext}$ as a bifurcation parameter, other parameters fixed. For very low value of $I_{\rm ext}$, a unique stable equilibrium point exists. Upon increasing $I_{\rm ext}$, the system changes stability through a saddle node bifurcation ${\rm SN}_{1}$ at $I_{\rm ext}\approx30.52$ and the unstable branch fold back via another saddle node bifurcation ${\rm SN}_{2}$ at $I_{\rm ext}\approx-39.57$. Between the two SN bifurcations, the system has three equilibria: one stable (lower branch) and two unstable (upper and middle branch), see Fig.~\ref{fig:Ibif15}. The upper unstable branch changes stability at a subcritical Hopf bifurcation ${\rm HB}$ at $I_{\rm ext}\approx6.656$ before the system returns to a rest state as $I_{\rm ext}$ increases. The unstable limit cycle emanated from ${\rm HB}$ fold back and changes to a stable limit cycle through a saddle node bifurcation of cycles ${\rm SNC}_{1}$ at $\text{I}_{\text{ext}}\approx 26.84$. The limit cycle loses stability at another ${\rm SNC}_{2}$ at $\text{I}_{\text{ext}}\approx 22.99$ before it terminates at $I_{ext}\approx 23.79$. %The onset of oscillations is through a Hopf bifurcation with nonzero frequency.% therefore for this parameter set the system behaviour is Type II excitability. 
  
\begin{figure}[htbp]
\centering
\begin{subfigure}[b]{.5\textwidth}
    \centering
    \caption{}
     \label{fig:Ibif3}
    \includegraphics[width =\textwidth]{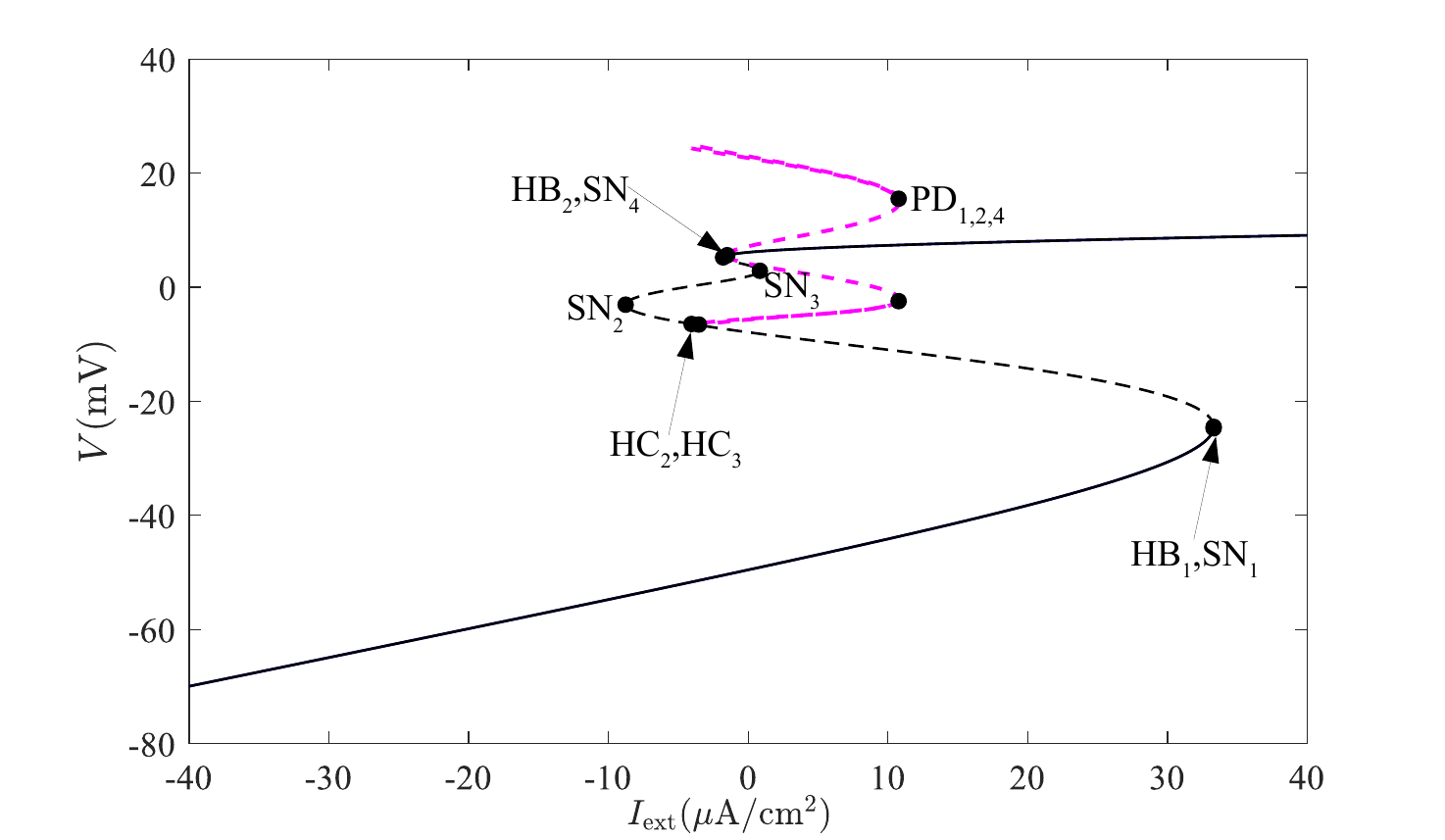}
  \end{subfigure}\\%
\begin{subfigure}[b]{.5\textwidth}
    \centering
    \caption{}
     \label{fig:Ibif3Z1}
    \includegraphics[width =\textwidth]{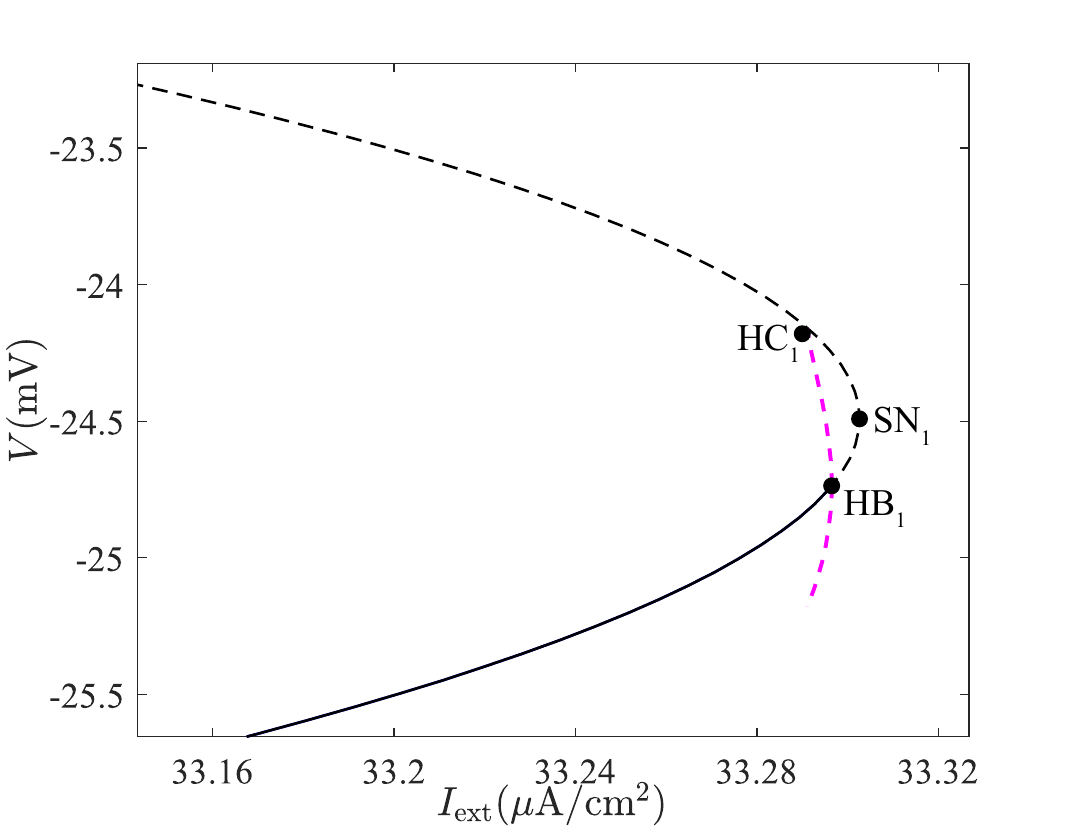}
  \end{subfigure}%
  \begin{subfigure}[b]{.5\textwidth}
    \centering
    \caption{}
    \label{fig:Ibif3Z2}
    \includegraphics[width =\textwidth]{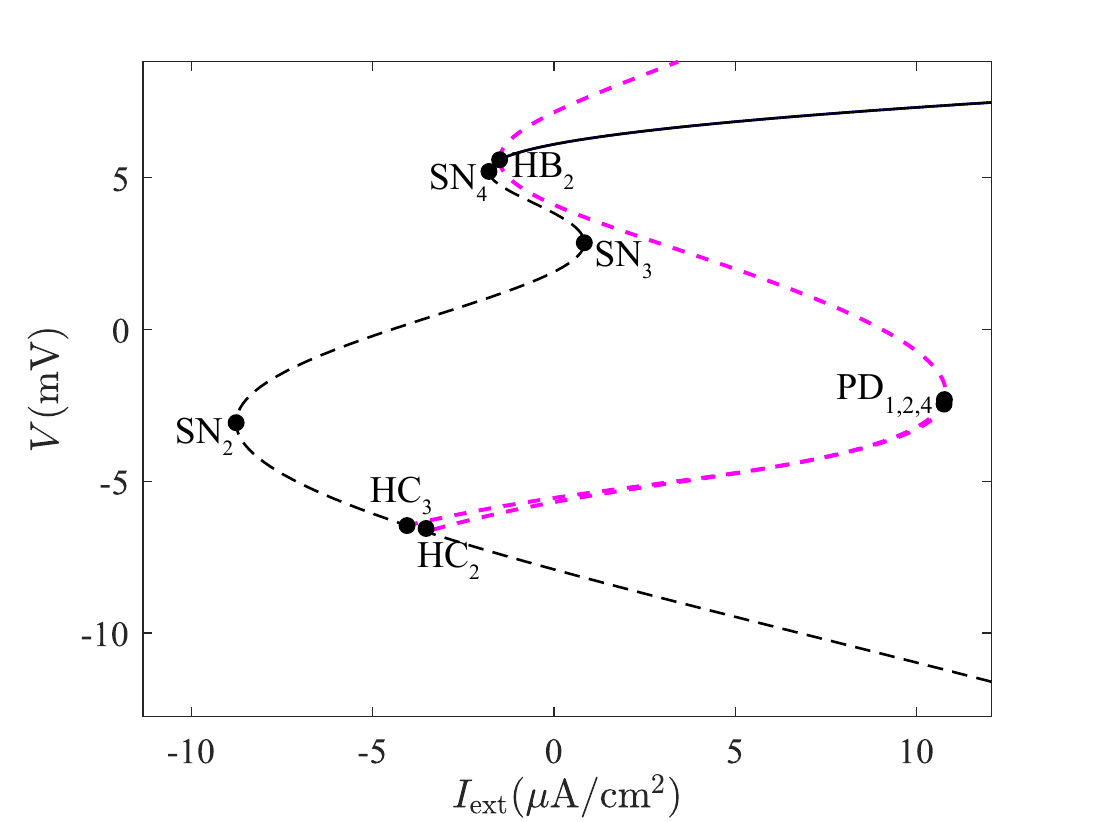}
  \end{subfigure}
 \caption{(a) Bifurcation diagram of membrane potential $V$ with $I_{\text{ext}}$ as a bifurcation parameter. (b)--(c) are enlargements of (a). and other parameters are fixed as in Sec.~\ref{sec:model}. The labels and other conventions are as in Fig.~\ref{fig:gnabifs}}
\label{fig:I_1_bifs}
\end{figure}

\begin{figure}[htbp]
\centering
  \begin{subfigure}[b]{.5\linewidth}
    \centering
    \caption{}
    \includegraphics[width=\textwidth]{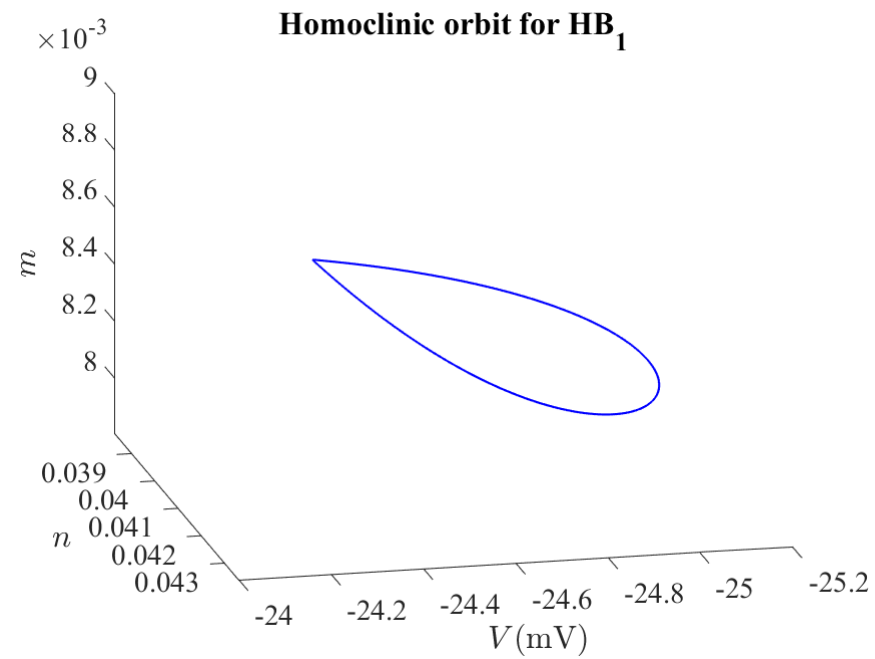}
    \label{fig:HB1-HOM}
  \end{subfigure}%  
  \begin{subfigure}[b]{.5\linewidth}
    \centering
    \caption{}
    \includegraphics[width=\textwidth]{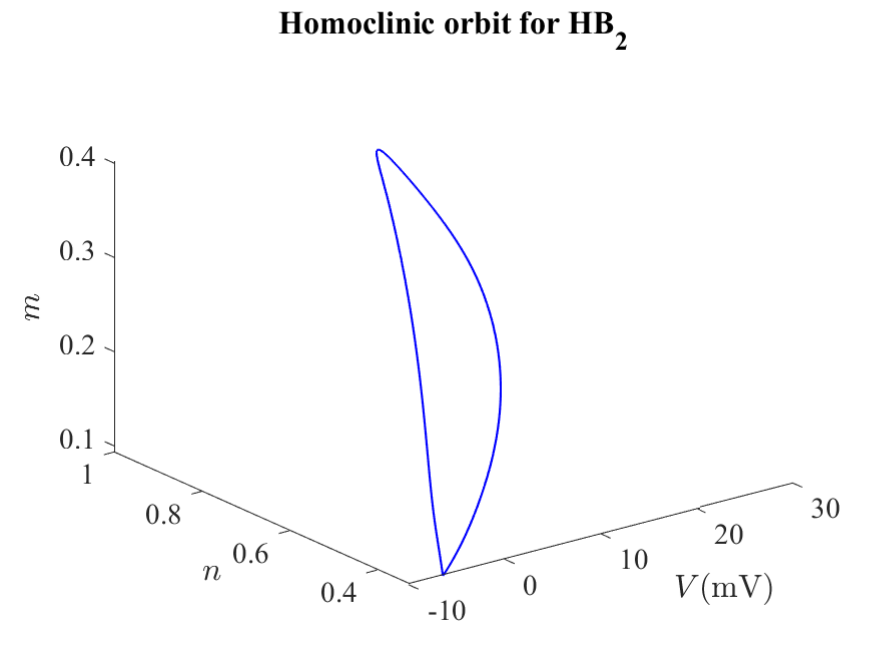}
    \label{fig:HB2-HOM}
  \end{subfigure}%  
  \caption{The curves of homoclinic orbits of the periodic oscillation emanated at (a) ${\rm HB}_{1}$; (b) ${\rm HB}_{2}$}
  \end{figure}

For set II, $v_6=3$ while other parameters are fixed as in Sec.~\ref{sec:model}. A bifurcation diagram of the membrane potential $V$ with $I_{\rm ext}$ as bifurcation parameter is shown in Fig.~\ref{fig:Ibif3}. For $I_{\rm ext}<-8.7715$, there exists a unique stable equilibrium point. Upon increasing $I_{\rm ext}$, the system loses stability through a subcritical Hopf bifurcation ${\rm HB}_{1}$ at $I_{\rm ext}\approx33.29650$. The unstable limit cycle emanated from ${\rm HB}_{1}$ ends in an homoclinic bifurcation ${\rm HC}_{1}$ at $I_{\rm ext}\approx33.2911$, see Fig.~\ref{fig:Ibif3Z1}. The curve of the homoclinic orbit is shown in Fig.~\ref{fig:HB1-HOM}. Increasing $I_{\rm ext}$ slightly there appears a saddle-node bifurcation ${\rm SN}_1$ at $I_{\rm ext}\approx33.2026$, the unstable branch fold back at another saddle-node bifurcation ${\rm SN}_2$ at $I_{\rm ext}\approx-8.7715$. 

As $I_{\rm ext}$ increases further, the system passes through another saddle node bifurcation ${\rm SN}_3$ at $I_{\rm ext}\approx0.8353$. For $I_{\rm ext}\in[{\rm SN}_{2},{\rm SN_{3}}]$, there exist three equilibria; one stable and two unstable. The branch of ${\rm SN}_{3}$ bifurcation folds at another saddle-node bifurcation $\text{SN}_{4}$ at $I_{\rm ext}\approx-1.7961$, and the unstable upper branch becomes stable in another subcritical Hopf bifurcation $\text{HB}_{2}$. For $I_{\rm ext}\in[{\rm SN}_{4},{\rm HB_{2}}]$, there exist five equilibria; one stable and four unstable equilibria. Also, for $I_{\rm ext}\in[{\rm HB}_{2},{\rm SN_{3}}]$, there exist five equilibria; two stable and three unstable. For this parameter values, the system is bistable, that is, coexistence of two stable equilibria. To the right of ${\rm SN}_{1}$, the system has a unique stable equilibrium.

\begin{figure}[htbp]
\centering
  \begin{subfigure}[b]{.5\linewidth}
    \centering
    \caption{}
    \includegraphics[width=\textwidth]{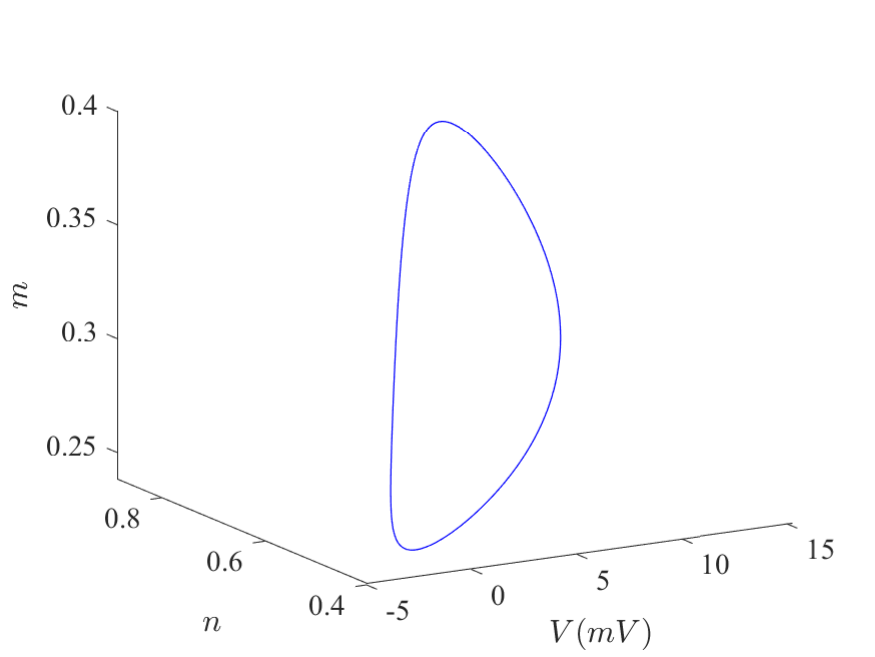}
    \label{fig:PD1-I}
  \end{subfigure}%  
  \begin{subfigure}[b]{.55\linewidth}
    \centering
    \caption{}
    \includegraphics[width=\textwidth]{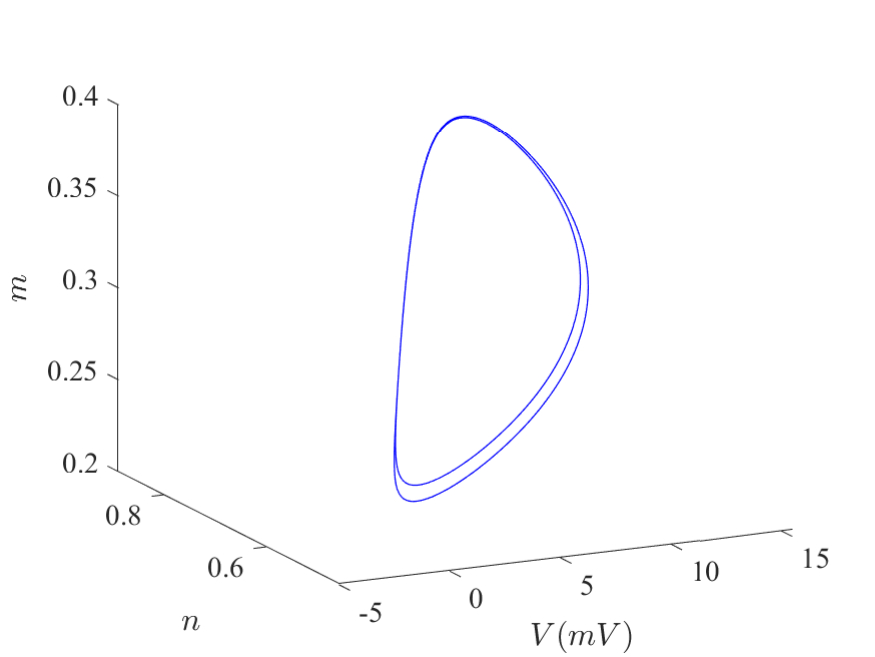}
    \label{fig:PD2-I}
  \end{subfigure}\\%  
  \begin{subfigure}[b]{.55\linewidth}
    \centering
    \caption{}
    \includegraphics[width=\textwidth]{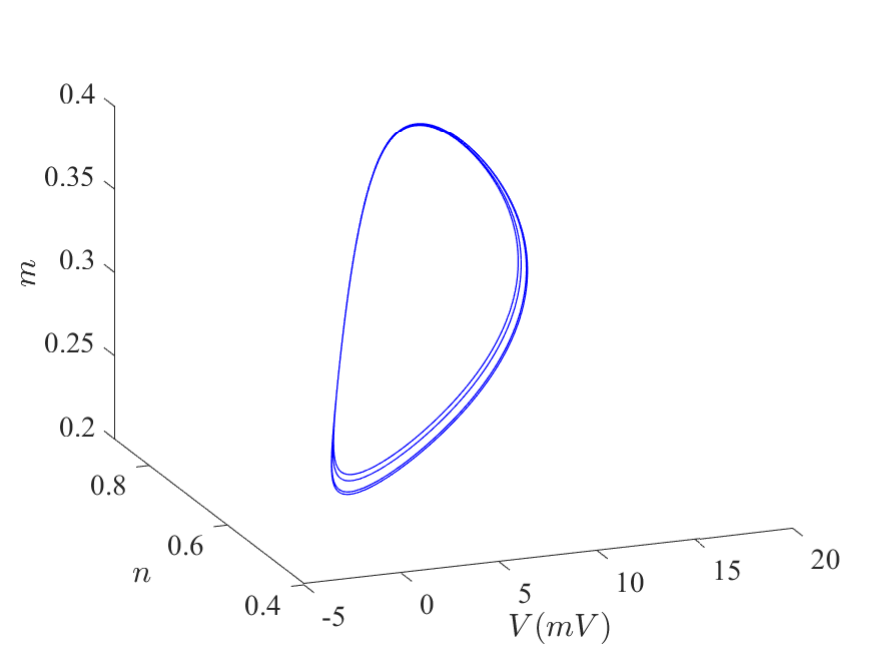}
    \label{fig:PD4-I}
  \end{subfigure}%  
  \caption{ Phase-space of \eqref{eq:1.1a}--\eqref{eq:4.1a} showing the period-doubling bifurcations in response to variation of $I_{ext}$. (a) Period-1 (b) Period-2 (c) Period-4, respectively}
  \label{fig:PD_I}
  \end{figure}

The unstable limit cycle generated at the Hopf bifurcation ${\rm HB}_{2}$ fold back at $I_{\rm ext}\approx10.80$ and slightly after the fold point appears a period-doubling bifurcation ${\rm PD}_1$ at $I_{\rm ext}\approx10.77$. At $\text{PD}_1$, the limit cycle bifurcates into unstable double-period and unstable limit cycles, and they both end in an homoclinic bifurcation, see Fig.~\ref{fig:Ibif3Z2}. The curve of the homoclinic orbit is shown in Fig.~\ref{fig:HB2-HOM}. Continuation from the period-doubling ${\rm PD}_1$ results in period-doubling bifurcation ${\rm PD}_{2}$, subsequently, the ${\rm PD}_{2}$ results in period-doubling bifurcation ${\rm PD}_{4}$. Table~\ref{Tab:PD2} shows the parameter values for the period-doubling and homoclinic bifurcations and their corresponding periods as $I_{\rm ext}$ is varied. The projections of periodic trajectories for period-1, 2, 4 onto $(V,n,m)$ phase space are shown in Fig.~\ref{fig:PD_I}.
 \begin{table}[htbp]
\centering
\caption{Summary of the parameter values and period of period doubling and homoclinic bifurcations that arise as $I_{\rm ext}$ is varied}
\label{Tab:PD2}       
\begin{tabular}{lll}
\hline\noalign{\smallskip}
Bifurcation point & \hspace{3mm} $I_\text{ext}$ & \hspace{1mm} Period \\
\noalign{\smallskip}\hline
\noalign{\smallskip}
${\rm PD}_{1}$ & 10.7705 &   33.5585\\[1mm]

${\rm PD}_{2}$ &  10.7584 &   67.1396\\[1mm]

${\rm PD}_{4}$ &  10.7555 &  134.353 \\[1mm]

${\rm HC}_{1}$&  33.2911 &  2.61499E+08\\[1mm]

${\rm HC}_{2}$ &  -4.05553 &  3.95045E+09 \\[1mm]
\noalign{\smallskip}\hline
\end{tabular}
\end{table}
\subsection{\textbf{Two Parameter Bifurcation Analysis}}
\begin{figure*}[htbp]
    \centering
    \includegraphics[scale=0.32]{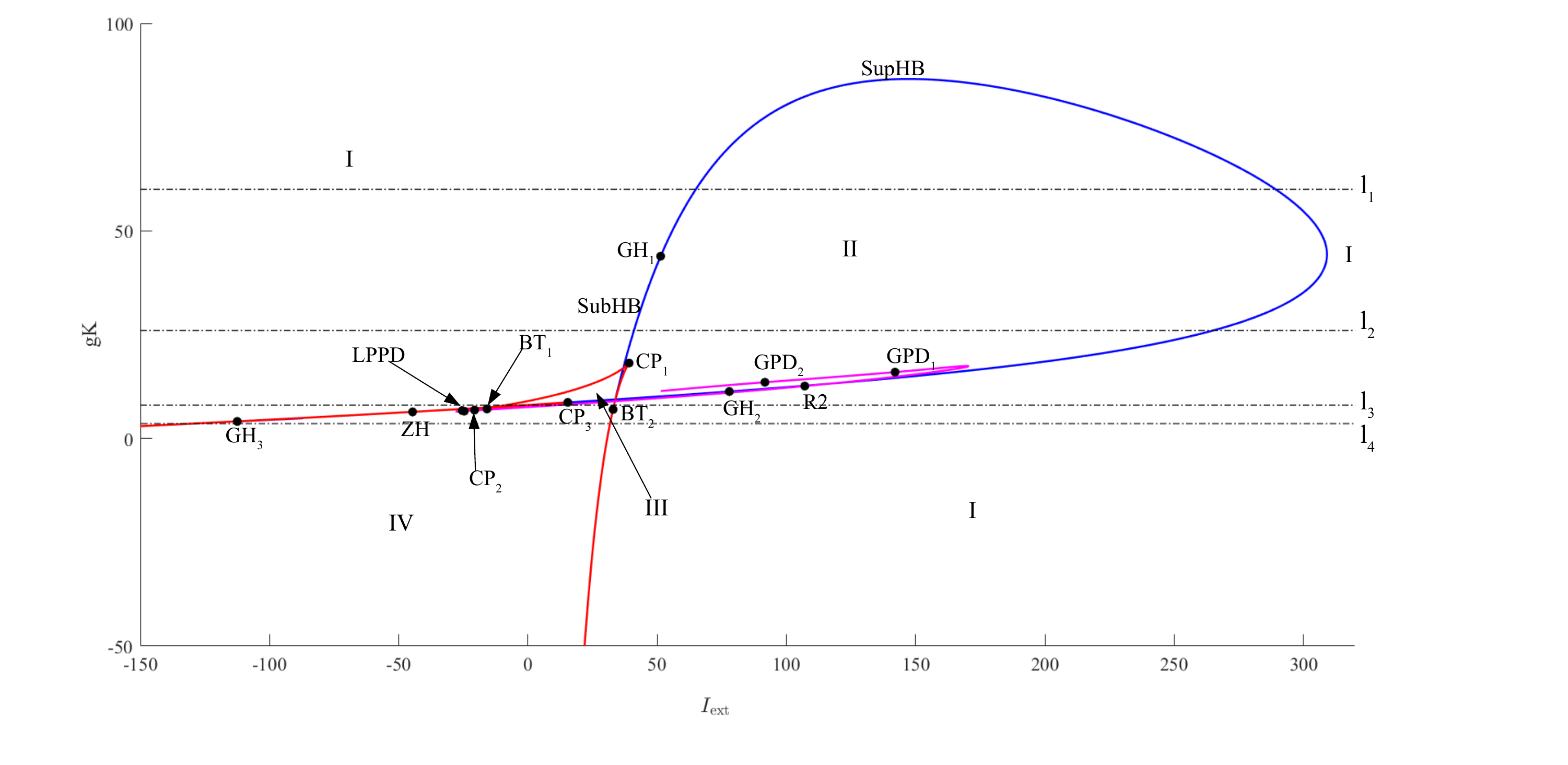}
    \caption{Two parameter bifurcation diagram of \eqref{eq:1.1a}--\eqref{eq:4.1a} in the $(I_{\rm ext},g_{\rm K})$-plane for parameter set II in Sect.~\ref{sec:Iext} and other parameter values as in Sect.~\ref{sec:model}. The values of $g_{\rm K}$ in ${\rm l}_{1}$, ${\rm l}_{2}$, ${\rm l}_{3}$, ${\rm l}_{4}$ are $60$, $26$, $8$ and $3.5$, respectively. The blue, red and magenta curves are the loci of Hopf bifurcation, saddle-node bifurcation, and period doubling bifurcation. The labels for the codimension-2 bifurcations are explained in Table~\ref{Tab:3}. The invariant sets that exist in each region are listed in Table~\ref{tab:regions}}
    \label{fig:2par}
\end{figure*}

In this section we perform two parameter bifurcation analysis of \eqref{eq:1.1a}--\eqref{eq:4.1a} in $(I_{\rm ext},g_{\rm K})$ plane. The bifurcation diagram shown in Fig.~\ref{fig:2par} is produced via numerical continuation software MATCONT \citep{Matcont2003}. The software implements Moore-Penrose continuation method to compute family and path of existing solution curves as parameters are varied. It is able to detect various kinds of bifurcations, switch to and compute the bifurcated branches, and allows us to follow the loci of the bifurcations in two parameters to detect codimension-2 bifurcation points. The step-by-step procedures for generating the codimension-2 bifurcation diagram Fig~\ref{fig:2par} in the GUI of MATCONT are given below:
\begin{enumerate}
\item[i.] First we integrate \eqref{eq:1.1a}--\eqref{eq:4.1a} from initial state variable values $(V,m,n,w)=(-20, 0, 0, 0)$ until the solution converges to an equilibrium point.
\item[ii.] Then we compute the equilibrium curve with $I_{\rm ext}$ as continuation parameter. To initialise the equilibrium continuation from the last point in (i), we set $I_{\rm ext}=50$, $\verb|ntst|=40$, and $\verb|ncol|=4$ in the \textbf{Starter} window and then compute \textbf{Forward} and \textbf{Backward}. Two Hopf bifurcations and four saddle-node bifurcations of equilibria are detected along the curve. The MATCONT window during the computation of the equilibrium curve is shown in Fig.~\ref{fig:matcontwindow}. 
\begin{figure*}[htbp]
    \centering
    \includegraphics[scale=0.3]{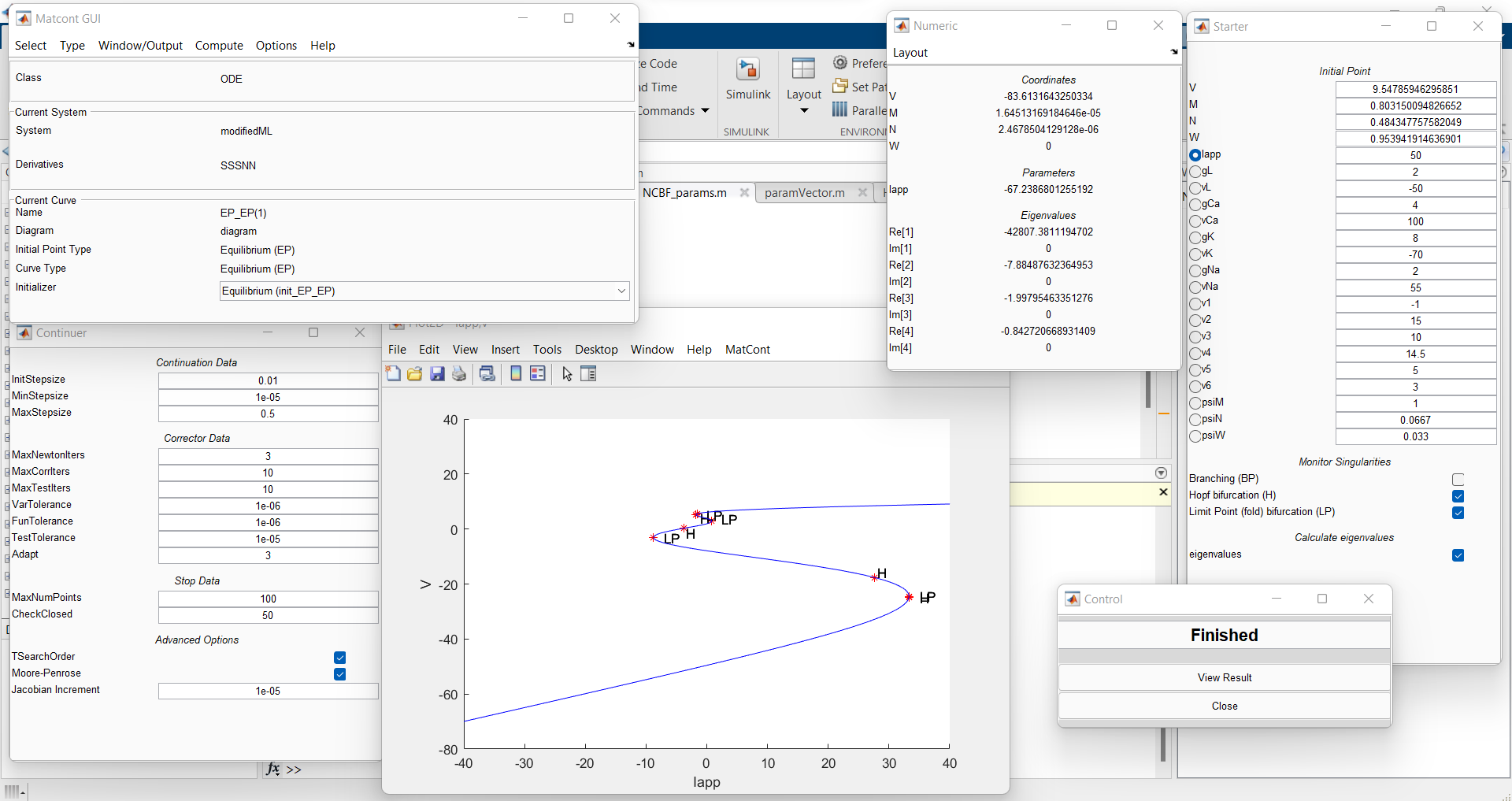}
    \caption{MATCONT window during the computation of the equilibrium curve}
    \label{fig:matcontwindow}
\end{figure*}
\item[iii.] Next we compute the limit cycles from the Hopf bifurcations. In the \textbf{Starter} window we set $I_{\rm ext}$ as bifurcation parameter and activate $\verb|period|$ to follow the period of oscillation along the continuation. We compute \textbf{Forward} to start the continuation from the Hopf bifurcation in the lower branch, MATCONT detects no special point except that the unstable limit cycle that emanates from the Hopf bifurcation  terminates at an homoclinic bifurcation, see Fig.~\ref{fig:PO_HB2}. Similarly, we compute \textbf{Forward} to start continuation from the Hopf bifurcation in the upper branch, an unstable limit cycle emanated from the Hopf bifurcation also terminated an homoclinci bifurcation and along the computation three period-doubling bifurcations are detected, see Fig.~\ref{fig:PO_HB1}.
\begin{figure}[htbp]
\centering
\begin{subfigure}[b]{.5\textwidth}
    \centering
    \caption{}
     \label{fig:PO_HB2}
    \includegraphics[width =\textwidth]{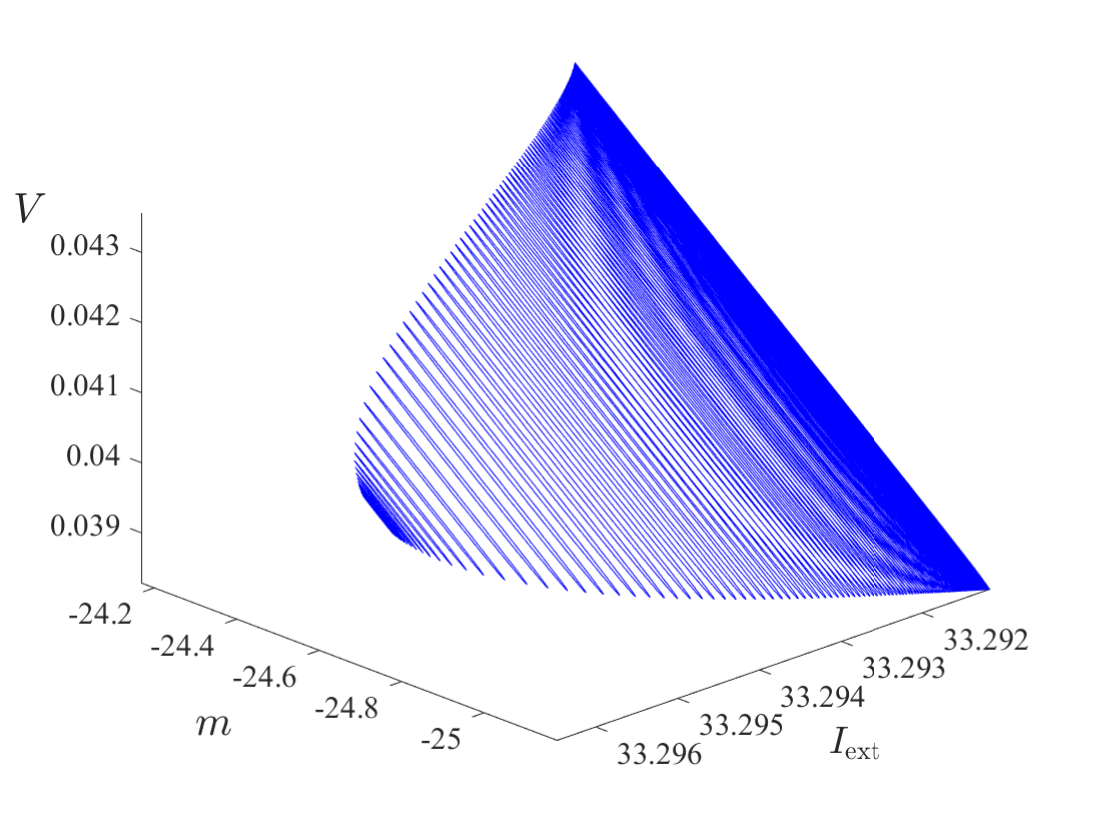}
  \end{subfigure}%
\begin{subfigure}[b]{.5\textwidth}
    \centering
    \caption{}
     \label{fig:PO_HB1}
    \includegraphics[width =\textwidth]{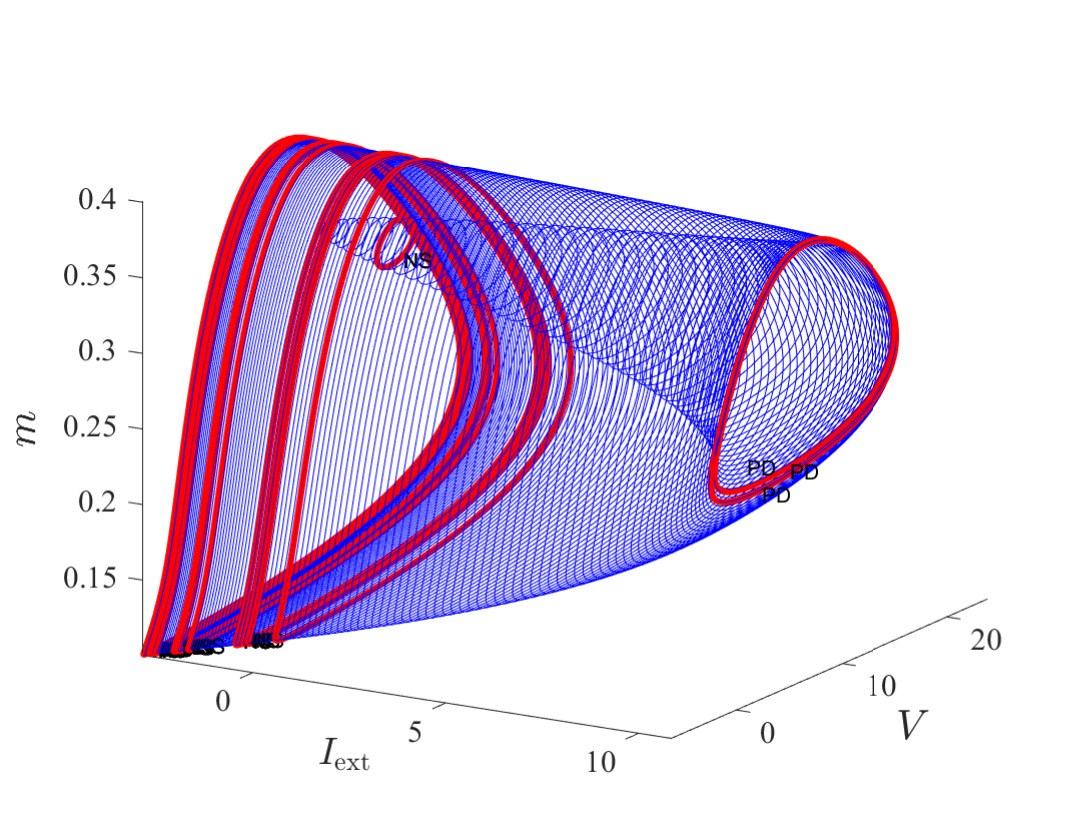}
  \end{subfigure}%
 \caption{A plot of the limit cycle that emanates from (a) the Hopf bifurcation in the lower branch; (b) the Hopf bifurcation in the upper branch of the equilibrium curve shown in Fig.~\ref{fig:matcontwindow}}
\label{fig:limitcycles}
\end{figure}

\item[iv.]  Finally, in the \textbf{Continuer} window we set $\verb|MaxStepSize|=1$ and select $I_{\rm ext}$ and $g_{\rm K}$ as bifurcation parameters in the \textbf{Starter} window. We then compute \textbf{Forward} and \textbf{Backward} at the Hopf bifurcation to produce the Hopf locus. Similarly, the loci of the saddle-node bifurcation and period-doubling bifurcation are initialised from each bifurcation points, respectively. Several codimension-2 bifurcations are detected and their descriptions are explained in Table.~\ref{Tab:3}. 
\begin{table}[htbp]
\centering
\caption{Abbreviations of codimension-two bifurcations}
\label{Tab:3}       
\begin{tabular}{ll}
\hline%\noalign{\scriptsizeskip}
Bifurcation & Abbreviation \\
\hline
%\noalign{\scriptsizeskip}\hline\noalign{\scriptsizeskip}
Cusp bifurcation & $\text{CP}_i$ \hspace{1.5mm}$i=1,2,3$ \\[1mm]
Bogdanov-Takens bifurcation & $\text{BT}_i$ \hspace{1.5mm}$i=1,2$  \\[1mm]
Generalized Hopf bifurcation & $\text{GH}_i$ \hspace{1.5mm}$i=1,2,3$ \\[1mm]
Zero-Hopf bifurcation & ZH \\[1mm]
Generalised Period Doubling bifurcation &  $\text{GPD}_i$ \hspace{1.5mm}$i=1,2$\\[1mm]
1:2 Resonance & $\text{R2}$ \\[1mm]
Flip-flop bifurcation & $\text{LPPD}$ \\[1mm]
\hline
%\noalign{\scriptsizeskip}\hline
\end{tabular}
\end{table}

Fig.~\ref{fig:2par} is divided into regions with respect to different types of dynamical behaviour and we have assigned each region a number, see Table~\ref{tab:regions}. In the remainder of this section we describe the dynamics of model \eqref{eq:1.1a}--\eqref{eq:4.1a} as $I_{\rm ext}$ and $g_{\rm K}$ are varied.
\end{enumerate}

\begin{figure}[htbp]
\centering
\begin{subfigure}[b]{.5\textwidth}
    \centering
    \caption{}
     \label{fig:gk60}
    \includegraphics[width =\textwidth]{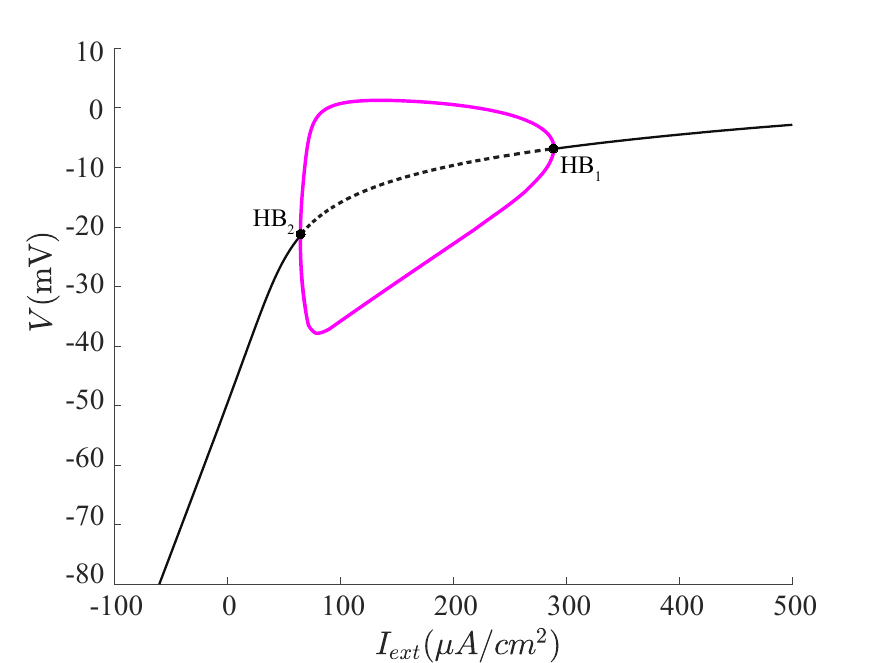}
  \end{subfigure}%
\begin{subfigure}[b]{.5\textwidth}
    \centering
    \caption{}
     \label{fig:gk26}
    \includegraphics[width =\textwidth]{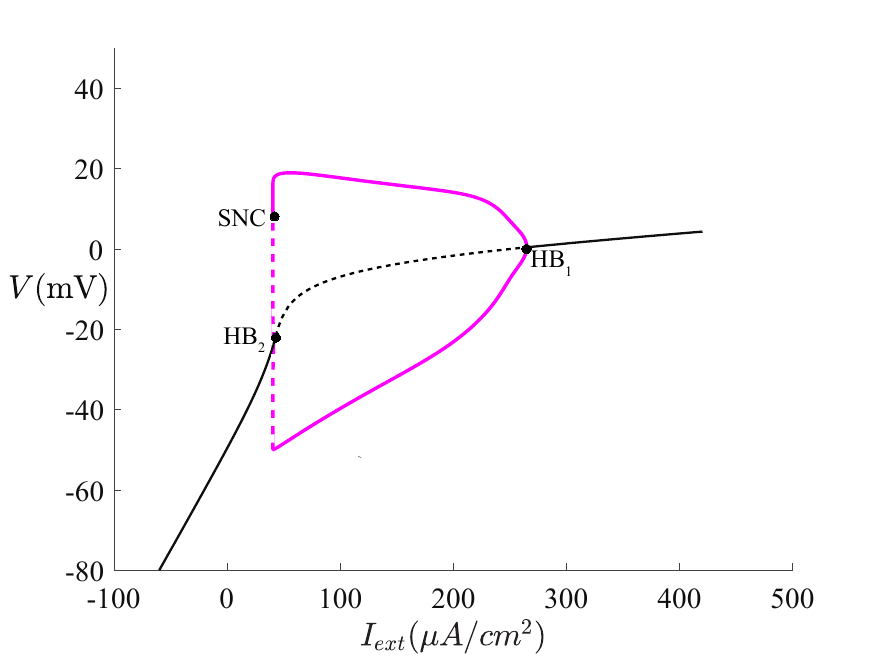}
  \end{subfigure}%
 \caption{(a) A codimension-1 bifurcation diagram along line ${\rm l}_{1}$ with $g_{\rm K}=60$. (b) A codimension-1 bifurcation diagram along line ${\rm l}_{2}$ with $g_{\rm K}=26$. The labels and other conventions are as in Fig.~\ref{fig:gnabifs}}
\label{fig:codim-1_gk}
\end{figure}

%\begin{landscape}
\begin{table}[htbp]
\centering
\caption{Summary of the six different combinations of equilibria and limit cycles that arise in Fig.\ref{fig:2par} and its magnifications,  Figs.~\ref{fig:zoom2par3}, \ref{fig:zoom2par}, and \ref{fig:zoom2par1}}
\label{tab:regions}       
\begin{tabular}{l|l}
\hline
Region & Existence of equilibria and limit cycles \\
\hline
& \\
 I  & One stable equilibrium, no limit cycles (rest state).\\[2mm]
\hline
&   \\
II & One unstable equilibrium, one stable limit cycle.\\[2mm]
\hline
& One stable equilibrium, two unstable equilibria,\\
III &   no limit cycles.\\[2mm]
\hline
&   Two stable equilibria, one unstable equilibrium,\\
IV    & no limit cycles. \\[2mm]
\hline
& One stable equilibrium, four unstable equilibria, \\
V  &  one unstable limit cycle.\\[2mm]
\hline
&   Two stable equilibria, three unstable equilibria,\\
VI  &  one unstable limit cycle.\\[2mm]
\hline
\end{tabular}
\end{table}
%\end{landscape}
For  sufficiently large values of $g_{\rm K}$, there are two supercritical Hopf bifurcations ${\rm HB}_{1}$ and ${\rm HB}_{2}$. Thus for slice ${\rm l}_{1}$ in Fig.~\ref{fig:2par} there are period solutions in region II. A codimension-1 bifurcation diagram along slice ${\rm l}_{1}$ for which $g_{\rm K}=60$ is shown in Fig.~\ref{fig:gk60}. The stable equilibrium solution loses stability through a Hopf bifurcation ${\rm HB}_{2}$ as ${\rm I}_{\rm ext}$ is varied. A stable limit cycle emanated from ${\rm HB}_{2}$ ends in another Hopf bifurcation ${\rm HB}_{1}$ before the equilibrium regains stability via ${\rm HB}_{1}$. Here the system passes through regions I$\to$II$\to$I. As the value of $g_{\rm K}$ decreases, there appears a generalised Hopf bifurcation, denoted ${\rm GH}_{1}$, on the Hopf bifurcation locus at $g_{\rm K}\approx 43.9007$. This is a codimension-2 point where the ${\rm HB}$ locus changes from supercritical SupHB to subcritical SubHB \citep{KuznetsovY.A.1995ElementsTheory}. Below the ${\rm GH}_{1}$, there are two Hopf bifurcations, a subcriticcal and a supercritical. Fig.~\ref{fig:gk26} is a bifurcation diagram along slice ${\rm l}_{2}$ in Fig.~\ref{fig:2par} for which $g_{\rm K}=26$. The system passes through regions I$\to$II$\to$I as in the previous case (slice ${\rm l}_{1}$) except that the stable equilibrium solution in region I loses stability through a subcritical Hopf bifurcation ${\rm HB}_{2}$. An unstable limit cycle emanated from ${\rm HB}_{2}$ changes stability via a saddle-node bifurcation of limit cycles (SNC), the stable limit cycle ends in a supercritical Hopf bifurcation ${\rm HB}_{1}$ then  to the left of ${\rm HB}_{1}$ the equilibrium solution regains stability.

\begin{figure}[htbp]
\centering
\begin{subfigure}[b]{.5\textwidth}
    \centering
    \caption{}
     \label{fig:zoom2par3}
    \includegraphics[width =\textwidth]{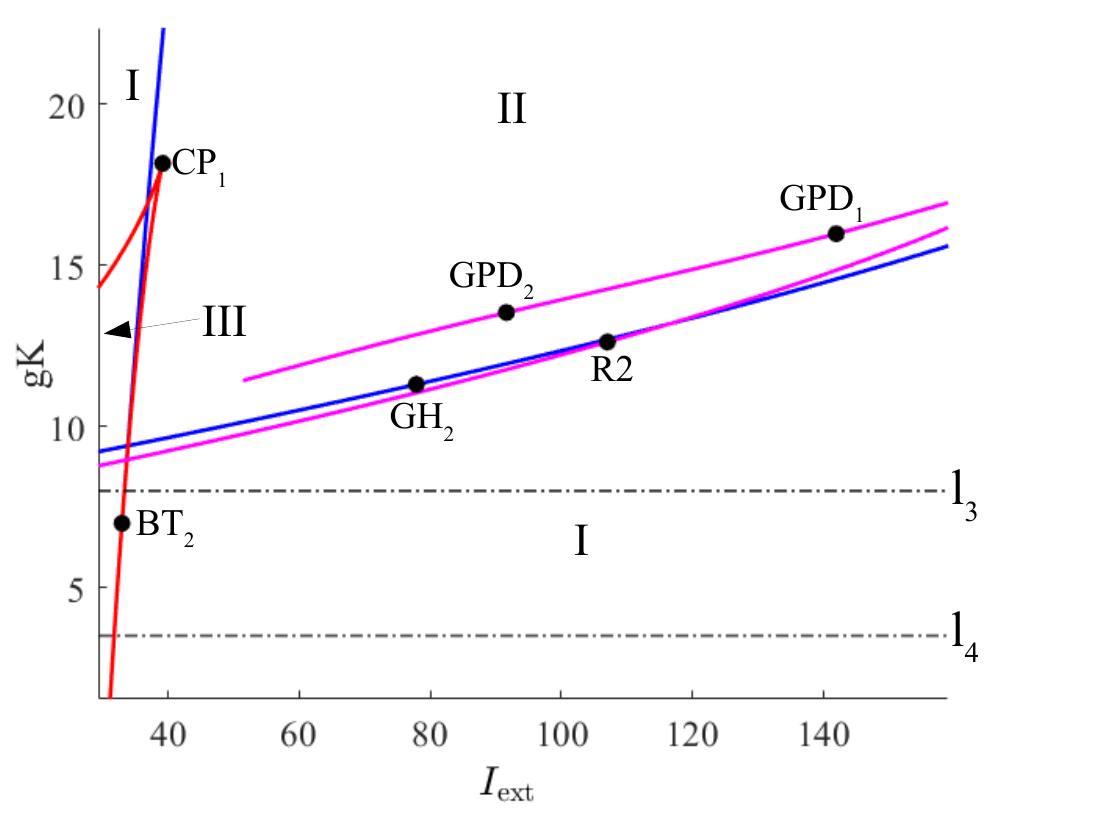}
  \end{subfigure}%
\begin{subfigure}[b]{.5\textwidth}
    \centering
    \caption{}
    \label{fig:zoom2par}
    \includegraphics[width =\textwidth]{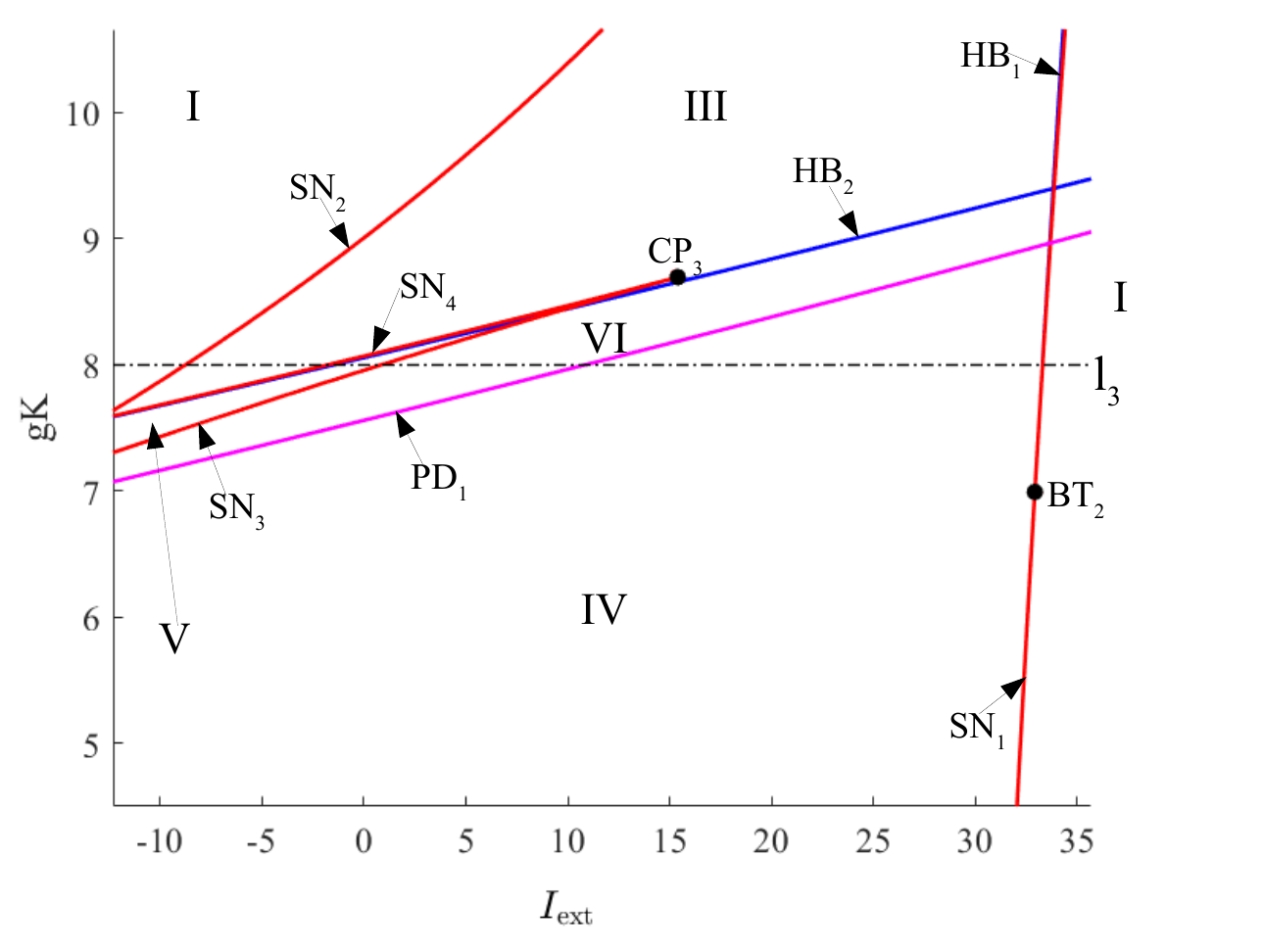}
  \end{subfigure}%
 \caption{(a)~An enlargement of Fig.~\ref{fig:2par}.~(b) An enlargement of Fig.~\ref{fig:2par} showing line ${\rm l}_{3}$ with $g_{\rm K}=8$. The labels and other conventions are as in Fig.~\ref{fig:2par} and Table.~\ref{Tab:3}}
\label{fig:zoom_2par}
\end{figure}

Upon further decrease in the value of $g_{\rm K}$, the loci of saddle-node bifurcations ${\rm SN}_{1}$ and ${\rm SN}_{2}$ collide and annihilate in a cusp bifurcation ${\rm CP}_{1}$ at $g_{\rm K}\approx18.1715$. As $g_{\rm K}$ decreases, a 1:2 resonance bifurcation ${\rm R2}$ and two generalised period-doubling bifurcations ${\rm GPD}_{1}$ and ${\rm GPD}_{2}$ appear on the locus of period doubling bifurcation at $g_{\rm K}\approx 12.624$, $15.982$, and $13.535$, respectively. Also, the loci of saddle-node bifurcations ${\rm SN}_{3}$ and ${\rm SN}_{4}$ collide and annihilate in a cusp bifurcation ${\rm CP}_{3}$ at $g_{\rm K}\approx8.6962$ and the supercritical Hopf bifurcation ${\rm SupHB}$ changes to subcritical Hopf bifurcation in another generalised Hopf bifurcation ${\rm GH}_{2}$ at $g_{\rm K}\approx 11.3037$, see Fig.~\ref{fig:zoom2par3}.

As the value of $g_{\rm K}$ is decreased below ${\rm CP}_{3}$, there exist four saddle-node bifurcations ${\rm SN}_{1}$, ${\rm SN}_{2}$, ${\rm SN}_{3}$ and ${\rm SN}_{4}$, an example is shown in Fig.~\ref{fig:zoom2par} along slice ${\rm l}_{3}$. The corresponding codimension-1 bifurcation diagram for which $g_{\rm K}=8$ is shown in Fig.~\ref{fig:Ibif3} and described in Sec.~\ref{sec:Iext}. The system passes through regions ${\rm I}\to{\rm III}\to{\rm V}\to{\rm VI}\to{\rm IV}\to {\rm I}$ in Fig.~\ref{fig:2par}. The loci of saddle-node bifurcations ${\rm SN}_{2}$ and ${\rm SN}_{3}$ collide and annihilate in a cusp bifurcation ${\rm CP}_{2}$ at $g_{\rm K}\approx18.1715$. As we decrease the value of $g_{\rm K}$ further, Bogdanov-Takens ${\rm BT}_{1}$ and ${\rm BT}_{2}$ occcur on the loci of saddle-nodes ${\rm SN}_{2}$ and ${\rm SN}_{1}$ at $g_{\rm K}\approx 7.1062$ and $g_{\rm K}\approx 6.9935$, respectively. The loci of subcritical Hopf bifurcations emanate from these codimension-2 points. These loci are tangential to ${\rm SN}_{2}$ and ${\rm SN}_{1}$ at these codimension-2 points. Observe also are zero-Hopf bifurcation ZH at $g_{\rm K}\approx 6.4099$, a codimension-2 where the locus of ${\rm HB}_{2}$ intersect the locus of ${\rm SN}_{4}$, and flip-flop bifurcation at $g_{\rm K}\approx 6.8379$ on the locus of period doubling bifurcation as $g_{\rm K}$ decreases. 

\begin{figure*}[htbp]
\centering
\begin{subfigure}{.5\textwidth}
  \centering
  \caption{}
  \includegraphics[width = \textwidth]{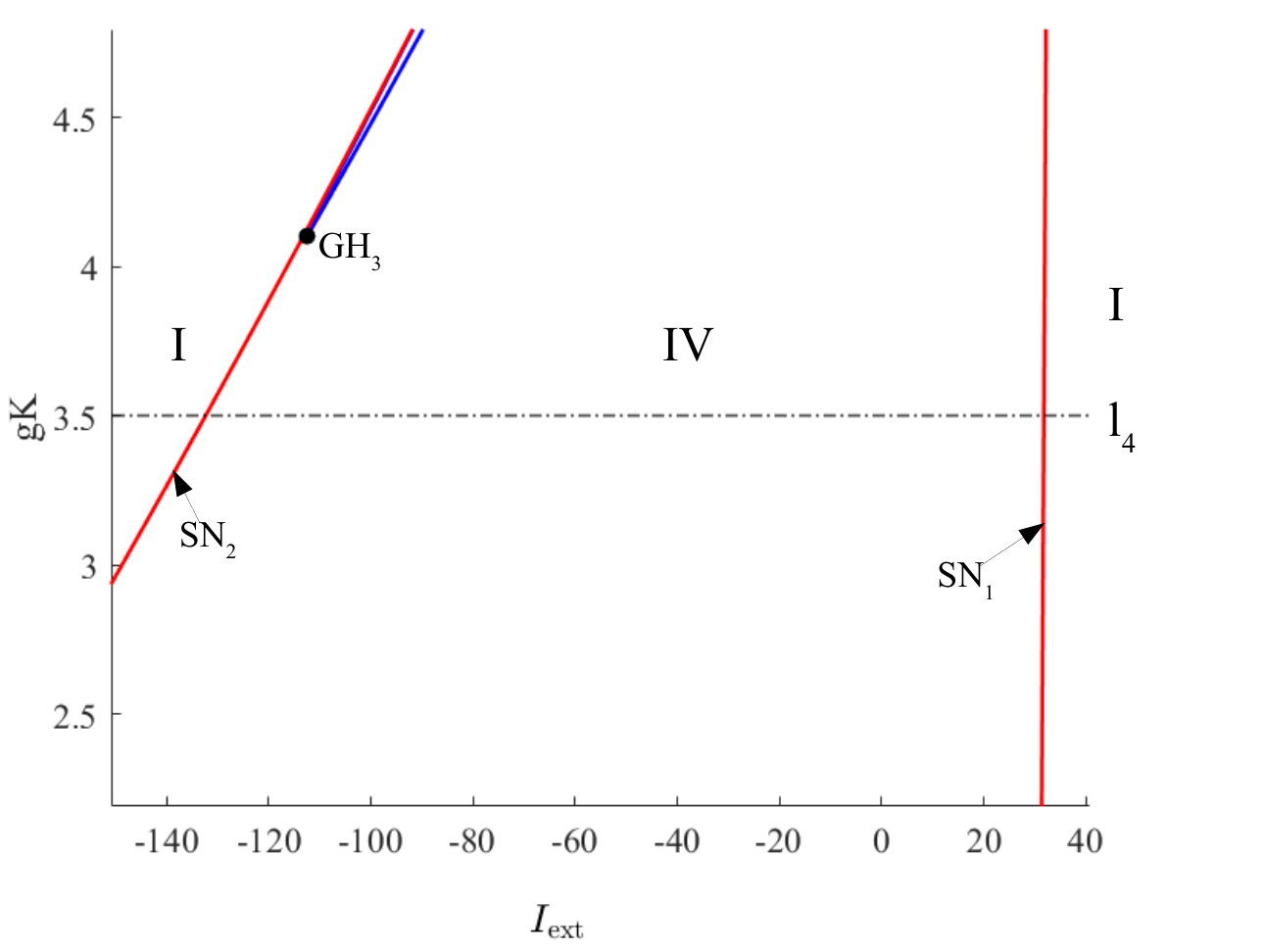}
    \label{fig:zoom2par1}
\end{subfigure}%
\begin{subfigure}{.5\textwidth}
  \centering
  \caption{}
  \includegraphics[width = \textwidth]{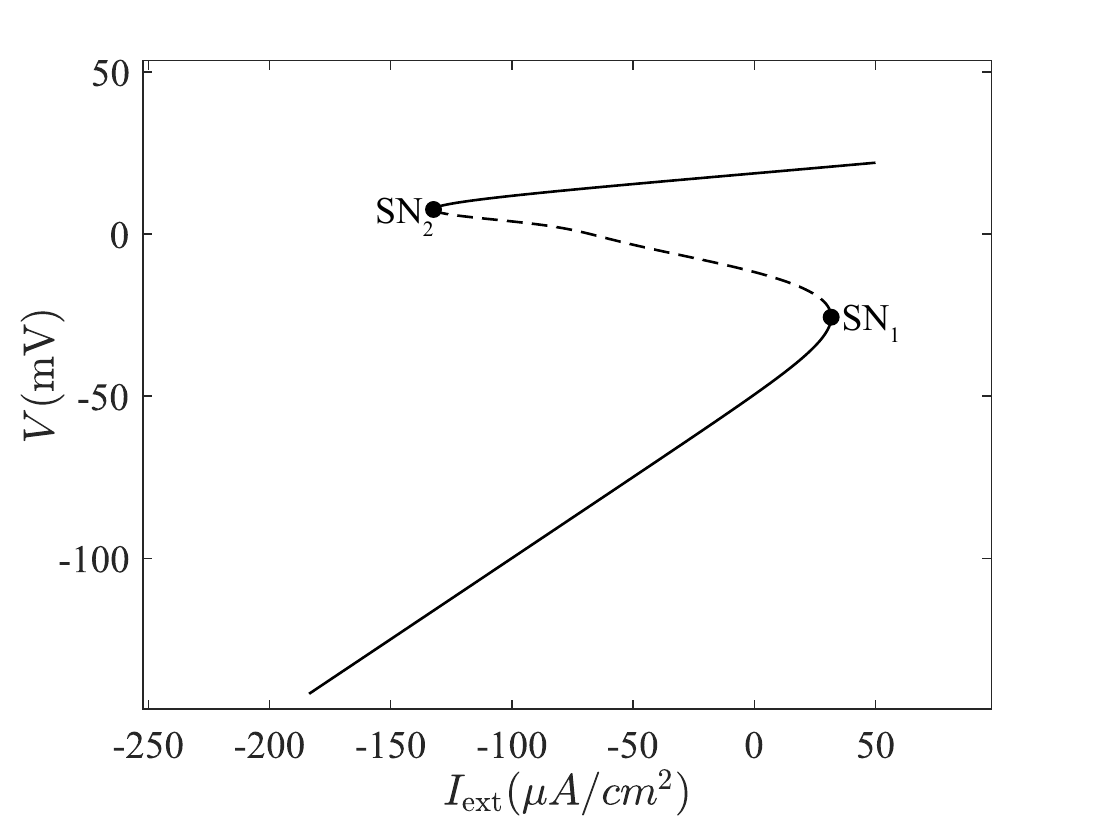}
    \label{fig:gk-3.5}
\end{subfigure}%
\caption{(a)~An enlargement of Fig.~\ref{fig:2par} showing lines ${\rm l}_{4}$.~(b)~A codimension-1 bifurcation diagram along line with $g_{\rm K}=3.5$. The labels and other conventions are as in Fig.~\ref{fig:2par} and Table.~\ref{Tab:3}}
\label{fig:enlarge2par}
\end{figure*}

Finally, as $g_{\rm K}$ is decreased further a generalised Hopf bifurcation, denoted ${\rm GH}_{3}$, occurs on the Hopf bifurcation locus ${\rm HB}_{2}$ at $g_{\rm K}\approx 4.1025$. Below this codimension-2 point, the only bifurcations that remain are the two saddle-node bifurcations ${\rm SN}_{1}$ and ${\rm SN}_{2}$. An example is shown Fig.~\ref{fig:zoom2par1} which is an enlargement of Fig.~\ref{fig:2par}. A bifurcation diagram along slice $l_{4}$ for which $g_{\rm K}=3.5$ is shown in Fig.~\ref{fig:gk-3.5}. Here the system passes through regions ${\rm I}\to{\rm IV}\to {\rm I}$.

\section{Conclusion}\label{sec:conclusion}
In this present paper, we have studied a 4D-ML model to explore the influence of second inward ${\rm Na}^{+}$ currents on electrical activities of excitable tissues. This work is motivated by the results in \citep{Ulyanova2018Voltage-dependentArterioles}, where it is reported that voltage-gated ${\rm Na}^{+}$ currents appear to contribute to the depolarising stage of action potentials in some excitable cells. We focused on addressing the influence of maximum conductances of ion channels on the dynamics of the membrane potential. Upon varying the conductance associated with the ${\rm Na}^{+}$ currents, $g_{\rm Na}$, the model exhibits different electrical activities.

With the aid of numerical bifurcation analysis, we examined the effects of parameters on the dynamical behaviour of the model. Our results showed that increasing the maximum conductance of sodium current $g_{\rm Na}$, the model transitions from rest state to periodic oscillations. For some values of $g_{\rm Na}$, the model shows complex behaviour, specifically, it undergoes cascades of period-doubling bifurcations. It was found that the bifurcation structure of varying the maximum conductance of potassium current $g_{\rm K}$ is qualitatively similar to that of varying the maximum conductance of calcium current $g_{\rm Ca}$ except in reverse. That is, increasing the value of $g_{\rm K}$ results in the same qualitative changes to the dynamics of the model as decreasing the value of $g_{\rm Ca}$.

We also showed qualitatively the effect of varying the external current $I_{\rm ext}$ on the dynamical behaviour of the model. Similar bifurcation diagram has been observed by \cite{Gall}, they discussed the bifurcation diagram in some detail, although without an explicit determination of the period oscillations thus their bifurcation diagram seems incomplete. However, in this work, we give a detailed bifurcation structure. We showed that the unstable periodic oscillations emanated from the two Hopf bifurcations terminate in homoclinic bifurcations. We also observed cascades of period-doubling PD bifurcations for some values of $I_{\rm ext}$. The existence of PD bifurcations is an indicator that the model can exhibit chaotic behaviour in some parameter regime. 

The codimension-2 bifurcation analysis in $(I_{\rm ext},g_{\rm K})$-plane gives further details on transitions between different electrical activities in the model. The electrical activities in the original ML model can be of Type I or II excitability depending on how the cell transitions from rest state to periodic oscillations is through a Hopf bifurcation.  \citep{hammed,Tsumoto2006BifurcationsModel}. In Type I excitability, the cell transitions from rest to an oscillatory state via a saddle-node on an invariant circle bifurcation and in Type II excitability the transition is via a Hopf bifurcation. In this work, the model exhibits only Type II excitability.

The results in this paper showed that the ${\rm Na}^{+}$ channels may influence the depolarisation stage of an action potential. It is hope that this model provides a framework that can aid in the understanding of various electrical activities in excitable cells. Based on the results of the present paper more complex behaviour is expected when two or more cells are coupled together, thus the dynamics of a network of cells would be addressed in future. The individual systems can be interconnected via ring-star network \citep{muni1} , two-dimensional lattice \citep{SHEPELEV2020}, multilayer network \citep{SHEPELEV2021} to account for various other spatio temporal patterns, chimera states.

%It is worth mentioning that in our analyses, the model is Type II excitability, that is, transition from rest state to periodic oscillation is through a Hopf bifurcation. Meanwhile, the original ML model can be of Type I or II excitability depending on how parameters are varied, and transitions between types of excitability have been studied extensively \citep{hammed,Tsumoto2006BifurcationsModel}. It will be of interest to check whether similar behaviour can be observed in this modified ML model, therefore, two-parameter bifurcation analysis could be considered as a future work.

%The high period confirms the homoclinic bifurcations.


\section*{Acknowledgment}
The authors are grateful for the extensive and constructive comments from the anonymous reviewers. SSM acknowledges Dr. Astero Provata for providing feedback, discussions on the manuscript and the School of Fundamental Sciences doctoral bursary funding during this research.





%

%\begin{acknowledgements}
 
%\end{acknowledgements}

\section*{Contributions}
The presented idea was conceived by HOF. He wrote the MATCONT and XPPAUT codes.
HOF and SSM carried out the numerical simulations and generated the figures. AA aided in the interpretation of the results. All authors jointly prepared the manuscript.

 \section*{Conflict of interest}
 The authors declare that they have no conflict of interest.
\bibliographystyle{spbasic}      % basic style, author-year citations

\bibliography{hamfat.bib}

\end{document}